\documentclass[12pt]{article}

\usepackage{arxiv}

\usepackage[utf8]{inputenc} 
\usepackage[T1]{fontenc}    
\usepackage{hyperref}       
\usepackage{url}            
\usepackage{booktabs}       
\usepackage{amsfonts}       
\usepackage{nicefrac}       
\usepackage{microtype}      
\usepackage{lipsum}
\usepackage{comment}
\RequirePackage{natbib}
\RequirePackage{amsthm,amsmath,amsfonts,amssymb}
\RequirePackage{graphicx}
\usepackage{placeins}

\title{To Deconvolve, or Not to Deconvolve: Inferences of Neuronal Activities using Calcium Imaging Data}

\author{
  Tong Shen\\
  Department of Statistics\\
  University of California\\
  CA, Irvine 92697\\

  \And 
  Gyorgy Lur\\
  Department of Neurobiology and Behavior\\
  University of California, Irvine.
  
  \And
    Xiangmin Xu\\
   Department of Anatomy and Neurobiology\\
   School of Medicine\\
   University of California\\
   Irvine, CA 92697
  
   \And
 Zhaoxia Yu \\
  Department of Statistics\\
  University of California\\
  CA, Irvine 92697\\
  \texttt{zhaoxia@ics.uci.edu} \\
}

\begin{document}
\maketitle

\begin{abstract}
With the increasing popularity of calcium imaging data in neuroscience research, methods for analyzing calcium trace data are critical to address various questions.  The observed calcium traces are either analyzed directly or deconvolved to spike trains to infer neuronal activities. When both approaches are applicable, it is unclear whether deconvolving calcium traces is a necessary step. In this article, we compare the performance of using calcium traces or their deconvolved spike trains for three common analyses: clustering, principal component analysis (PCA), and population decoding. Our simulations and applications to real data suggest that the estimated spike data outperform calcium trace data for both clustering and PCA. Although calcium trace data show higher predictability than spike data at each time point, spike history or cumulative spike counts is comparable to or better than calcium traces in population decoding. 
\end{abstract}


\section{Introduction}
With the technical advances in multiple fields \citep{dombeck2010functional, ghosh2011miniaturized, grienberger2012imaging, chen2013ultrasensitive, yang2017vivo}, calcium imaging has been increasingly adopted as a supplement or substitute to the traditional electrophysiological methods for measuring neuronal firing activities. Compared to electrophysiological methods, imaging methods offer flexibility in a number of ways such as better spatial resolutions, longer follow up time, ability to study not only awake but also freely moving animals, and the larger number of neurons that can be simultaneously measured. One trade-off for the greater flexibility is the requirement of more sophisticated pre-possessing, as the measurement of neuronal activities from calcium imaging is indirect, and its estimate of calcium concentration is complicated by several factors such as measurement noises and contamination of signals from non-neuronal cells \cite{johnston2020robust}. As a consequence, each calcium trace is only a proxy of the underlying spiking activities with a reduced signal-to-noise ratio and temporal resolution. 
Thus, recent comparisons of electrophysiology and calcium imaging data are timely to guide us on how to interpret the results obtained from calcium imaging data. Using matched neuron populations and experimental conditions, \cite{wei2020comparison} reported both consistent and divergent results between electrophysiology and calcium data on temporal dynamics (within each trial), trial-type selectivity, sources of variances, and population decoding. 


Despite the continued improvement in the quality of calcium imaging due to state of the art optical imaging devices and techniques, sensitive genetically encoded indicators, and pre-processing methods, extracting the underlying spike activities that would otherwise be accurately measured by electrophysiology data is still one major challenge in analyzing calcium imaging data in neuroscience research. The measured fluorescence intensity of calcium concentration is characterized by a rapid rise but slow decay after the occurrence of an action potential. As a result, numerous spike deconvolution methods have been proposed, from the simple thresholding with 2-3 standard deviations away from the calcium trace baseline to formal and fully Bayesian models \citep{yaksi2006reconstruction, vogelstein2010fast, pnevmatikakis2016simultaneous, jewell2018exact, pachitariu2018robustness, pnevmatikakis2019analysis}, including two of our recent work on
multi-trial data \citep{johnston2020robust, tong2021}.  Comparisons between estimated spike data and the true spike data from simulations or benchmark data usually showed that deconvolution leads to satisfactory results; on the other hand, the estimated spikes might be inconsistent across methods, which produces different estimates of firing rate, number of tuned neurons, and distribution of estimated firing rates \citep{evans2019use}. 

Some research problems, such as investigating temporal coding \citep{abeles1993spatiotemporal, abbott1994decoding, theunissen1995temporal} and the estimation of instantaneous firing rates, requires the precise timing of action potentials. In addressing many other scientific questions, however, either calcium traces or estimated spike data can be used. For visualization purposes, both heat maps of calcium traces and raster plots of estimated spike trains are commonly presented. Cluster analysis is another example. It has been conducted to group neurons or trials with similar temporal patterns using spike train data \citep{adler2012temporal, humphries2011spike}. Due to the increased availability of calcium recordings, in recent work, calcium trace data have also been directly used for clustering \citep{ozden2008identification, dombeck2009functional, barbera2016spatially}. Cluster analysis based on imaging data suggest that spatially compact neural clusters exist in awake mouse motor cortex at not only macrocircuitry but also microcircuitry levels \citep{dombeck2009functional}. These clusters may represent cell assemblies that are stable over days; their dynamics often represent unique behavioral states and carry useful encoding information for behaviors \citep{barbera2016spatially}. When calcium imaging data are recorded, one can conduct cluster analysis using either calcium traces or the deconvolved spike data \citep{romano2017integrated}. In this situation, one natural question is should one conduct cluster analysis on calcium traces or the deconvolved data?

A related analysis is principal component analysis (PCA). Compared to cluster analysis, which groups observations to homogeneous clusters, PCA aims to extract component (features) that can keep as much as possible the variation in the original data. In neuroscience, PCA is frequently performed for dimension reduction and data visualization. Quantifying the dimensionality defined based on the corresponding eigenvalues may also shed light on the dynamics of the underlying neural circuits of various tasks and stimuli \citep{gao2017theory}. 
PCA analysis can be conducted to both electrically recorded spikes trains and optimally recorded calcium imaging data \citep{cunningham2014dimensionality}. One advantage of calcium recording over electrophysiology recording is that calcium imaging allows a large number of neurons to be recorded simultaneously. A popular method to compactly visualize the neuronal dynamics is to plot PCA trajectories over time using the first two or three components \citep{churchland2012neural, cunningham2014dimensionality}. 
PCA is more straightforward for calcium traces, which are continuous values. Methods aiming to identify components using spike train data often filtered such as Gaussian kernel \citep{churchland2012neural} and other choices \citep{paiva2010inner}. When the components of the true spike activities are of interest, PCA using calcium trace data might not be desirable, as a substantial source of the variation might be from the noises, rather than the signals, in the calcium trace data. In this situation, deconvolution might be helpful to recover the true underlying components. 

Another important analysis in neuroscience is decoding, which refers to finding the mapping from neural activities to either external stimuli such as presented images and experimental types or animal behavioral outcomes such as movements, speed, and positions, and decision making. An interesting question is whether and how information is represented in an ensemble of neurons. Thus, population decoding methods have been widely adopted to investigate of the joint activities of a group of neurons using multiple spike trains \citep{brown2004multiple}. For example, \cite{yang2020differential} analyzed simultaneously measured spike train data from two brain regions to compare their population decoding of choices in a two-alternative choice task. With the increased popularity of using calcium imaging to measure a large population of neurons, more and more population decoding analyses are conducted using calcium trace data. In most published work, the calcium trace data were first deconvolved to spike train data before they were used to population decoding. A recent study \citep{wei2020comparison} found that, as expected, the decoding accuracy of the deconvolved spike data from calcium traces was much lower than the eletrophysiology data; but counter-intuitively, calcium trace data have higher predictability than their deconvolved spike train data and the less noisy eletrophysiology data. A possible explanation suggested by \cite{wei2020comparison} is, due to the slow decay rate of the observed calcium transients, the calcium trace data provides prediction based on integrating effects rather than instantaneous decoding. This is evidenced by the improved predictability of the electrophysiology data when a filter of 1 second was applied. It is unknown whether it is beneficial to conduct deconvolution when ``integrating effects'' are taken consideration to make a fair comparison between calcium traces and their deconvolved spike trains.

In this paper, our goal is to examine the necessity of spike estimation for calcium trace data in three widely used methods, which are cluster analysis, PCA analysis, and population decoding.




\section{Cluster Analysis}
K-means is perhaps the easiest to understand but also the most widely used clustering method. The idea is to allocate a neuron to the cluster with the nearest centroid. Its improved versions have also been used in clustering either neurons or trials. For example, the meta k-means, which is a consensus clustering methods that aggregates clustering results from multiple runs of k-means, has been developed to increase the clustering stability of neurons \citep{ozden2008identification}. To account for clustering uncertainty when clustering trials, \cite{dunn1973fuzzy,bezdek2013pattern} introduced fuzzy k-means, a method that produces probabilities of cluster assignments to each trial. When comparing the impact of calcium trace and estimated spike data on cluster analysis, we choose the fuzzy k-means to take the inherent uncertainty in clustering into consideration. 

\subsection{Results based on simulated data}
We first compare the clustering accuracy based on calcium trace and the deconvolved spike data using a simulation study, where the true underlying cluster structure of the neurons are unknown. The spike train data we analyze here is a subset of the spike trains simulated by \cite{fellous2004discovering}. Each simulated data set consists of three clusters, which are treated as three neuron clusters here. The 35 neurons (spike trains) within each cluster share 4-6 spikes with the spike times uniformly distributed between the time interval (0-1 second). The following three ways are used to add noises at various levels. 
\begin{itemize}
    \item $15\%$ spike events are dropped randomly.
    \item $X$ extra random spikes are added to each train.
    \item All the spike times were jittered by a value drawn from a normal distribution with mean 0 and standard deviation $J$ million second (ms). 
\end{itemize}
We consider the following seven noise levels, from well separated clusters to very ``fuzzy'' cluster memberships. 

\begin{table}[h!]
\centering
 \begin{tabular}{|c c c c c c c c|} 
 \hline
 noise level & 2 & 3 & 4 & 5 & 6 & 7 & 8 \\ [0.5ex] 
 \hline
 $X$ & 2 & 3 & 4 & 8 & 11 & 15 & 20   \\ 
 $G$ & 1 & 3& 5& 10 & 15 & 20 &30 \\ 
 \hline
 \end{tabular}
\caption{The noise levels in the simulated spike trains from \cite{fellous2004discovering}. $X$ is the extra number spikes added per spike train; $G$ is the standard deviation when jittering the spike times. The labels of the noise levels from \cite{fellous2004discovering} are used.}
\label{table:1}
\end{table}

\begin{figure}[h]
		\centering
		\begin{tabular}{c c}
			\includegraphics[scale=0.76]{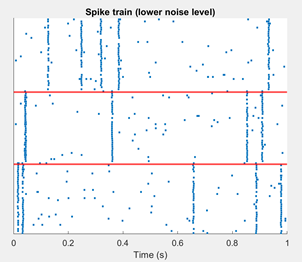} &
			\includegraphics[scale=0.77]{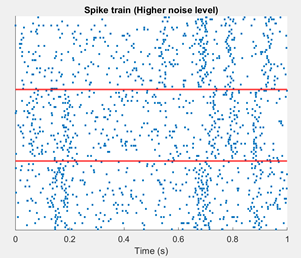}
		\end{tabular}
		\caption{Rasterplot of a simulated spike data with $G=3$ groups (separated by red lines) and 35 spike trains in each cluster. Left: with noise $X=3$ extra spikes and $J=3$ ms jitter. Right: with noise $X=11$ extra spikes and $J=20$ ms jitter. }
		\label{fig: sim example}
	\end{figure}

The raster plots for Figure \ref{fig: sim example} shows a set of spikes trains with low noise (left panel: level 2) and a set with high noise (right panel: level 8), respectively. In \cite{fellous2004discovering}, 30 data sets were generated for each of the seven noise levels to account for variations in data simulations. For each of the $30\times7=210$ data sets, we generate 100 sets of fluorescence traces, where each fluorescence trace is generated for a given spike train using an AR(1) model \citep{vogelstein2010fast}. In the AR(1) model, the calcium fluorescence for a neuron at a single trial $y(t), t=1,\cdots,T$ is modeled using the first-order auto-regressive model

\begin{equation*}
\begin{split}
     y(t) &= c(t)+\epsilon(t), \>\>\> \epsilon(t)\sim N(0,\sigma^2), \\
     c(t) &= \gamma c(t-1)+s(t), 
     \end{split}
\label{eqn:ar1}
\end{equation*}
where $c(t)$ denotes the underlying true calcium concentration, $s(t)$ represents the change in calcium concentration between time points $t-1$ and $t$ with $s(t)>0$ indicates a spike at time $t$ and $s(t)=0$ otherwise, $\gamma$ is the decay rate of calcium transients, and $\sigma^2$ denote the variance of the noise in measuring calcium concentration. In our simulations, the parameters we choose are as follows: the rate of decay $\gamma=0.96$, the magnitude of each spike $s(t)=1$ for any $s(t)>0$, and two levels of Gaussian noise $\sigma=0.1$ or $\sigma=0.3$. To estimate spikes from simulated calcium data, we use the $\ell_0$ penalized approach of \cite{jewell2018exact} for each calcium trace.

\begin{figure}[h]
		\centering
		\begin{tabular}{c c}
			\includegraphics[scale=0.45]{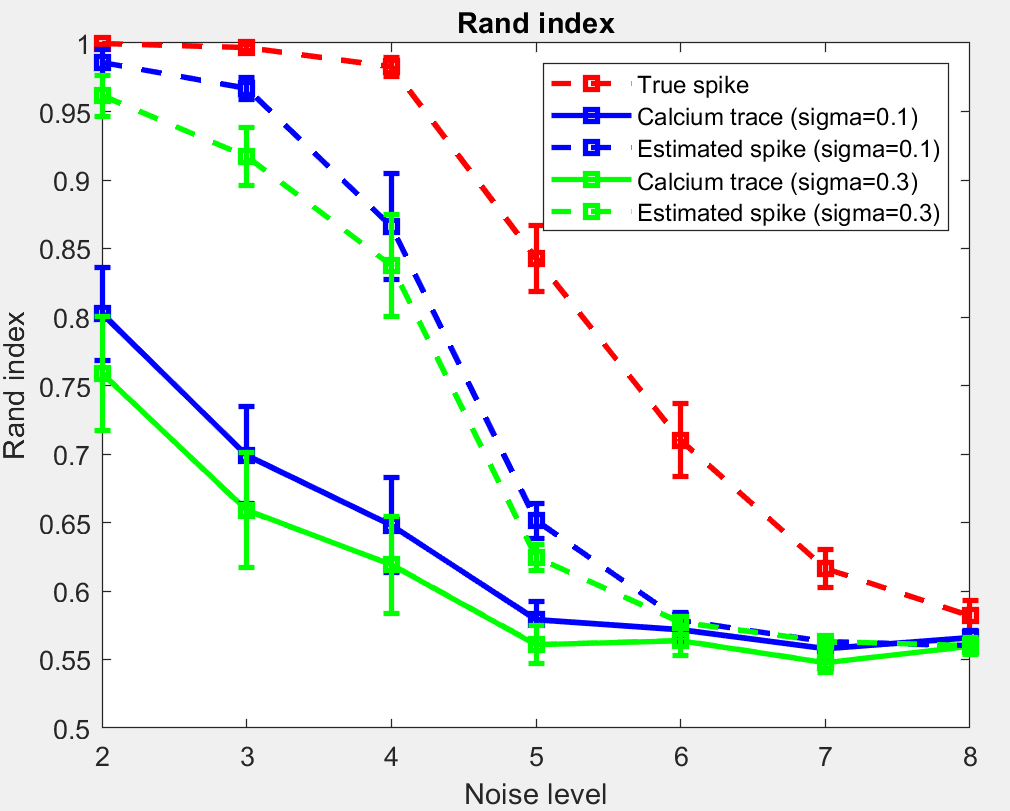}&\includegraphics[scale=0.45]{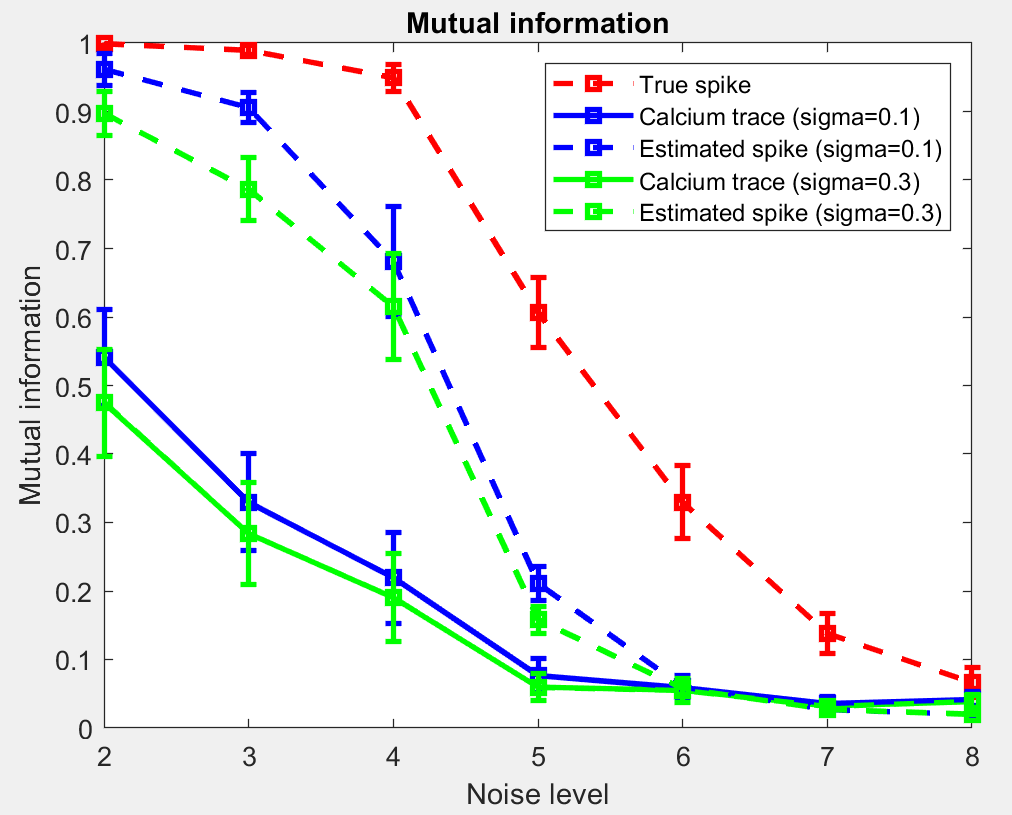}
		\end{tabular}
		\caption{Clustering results for simulated data over seven noise levels. Presented are means and 95\% confidence intervals over 30 data sets in each simulation setting. For the true spike data, each data point is averaged over 30 data sets; for data involving simulated calcium traces, at each noise level, clustering results from the 100 simulations for each of the 30 data sets are averaged first, and the results of the 30 data sets are then used to compute means and standard errors. Left: rand index. Right: mutual index. }
		\label{fig: sim_3clus}
	\end{figure}

We perform cluster analysis using the fuzzy k-means \citep{fellous2004discovering} with the assumption of three clusters to all the three types of data, namely, the calcium traces, the estimated spike trains from calcium traces, and the true spike trains. 
Their clustering accuracy is then quantified using two metrics - the rand index \citep{rand1971objective} and the normalized mutual information $I(A,B)$ \citep{danon2005comparing} between the estimated and true cluster memberships. Both metrics measure the consistency between two categorical variables, with the maximum $1$ indicating perfect agreement. Not surprising, as presented in Figure \ref{fig: sim_3clus}, the true spike data have the best clustering performance at all noise levels. Importantly, estimated spikes perform better than calcium trace at all noise levels.

In practice, the true number of clusters is unknown. To assess whether the results are sensitive to the choice of number of clusters, we also calculate the rand indices and mutual information for 4 and 5 clusters. The results (See appendix A.1) lead to the same conclusion, i.e., estimate spike data perform uniformly better than calcium trace on clustering.


\FloatBarrier

\subsection{A delayed response task}
\label{subsection:water_lick}
Next, we compare calcium traces and estimated spikes on clustering using calcium imaging data from two mice task studies. This application falls into the situation of no ground truth, as neither the true number of clusters nor the clustering membership is unknown a priori. This is a common situation, as benchmark data is not always available. Similar to \cite{pachitariu2018robustness}, who used concordance of between repeated trials estimation as a metric for spike detection, we will use rand index to examine the between-trial consistency for estimated clusters.  

In this multi-trial experiment, mice were first trained to discriminate pole location using their whiskers and reported the perceived pole position by licking \citep{li2015motor}. Their neuronal activities in the left anterolateral motor cortex region were then measured using two-photon calcium imaging using GCaMP6s. Each trial consisted of three epochs: sample epoch (mice presented with a vertical pole), delay epoch (the pole was removed), and response epoch (mice cued to give a response). Here we clustered the 67 neurons of a mouse that made 31 ``correct lick left'' trials and 21 ``correct lick right'' trials. Specifically, for each trial, we clustered the 67 calcium traces or the estimated spike trains into 4 clustering using the fuzzy k-means \citep{fellous2004discovering}. The rand index between every pair of two trials was calculated to assess the trial-to-trial consistencies. 

\begin{figure}[h]
		\centering
		\begin{tabular}{c c}
			\includegraphics[scale=0.47]{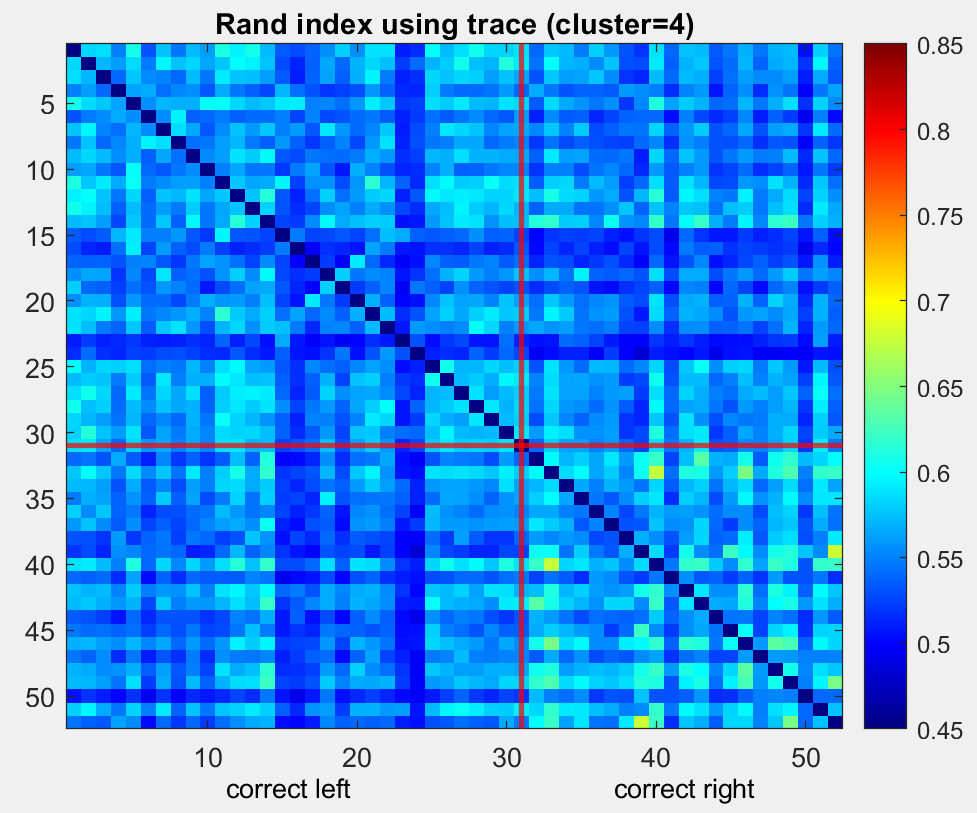}&\includegraphics[scale=0.5]{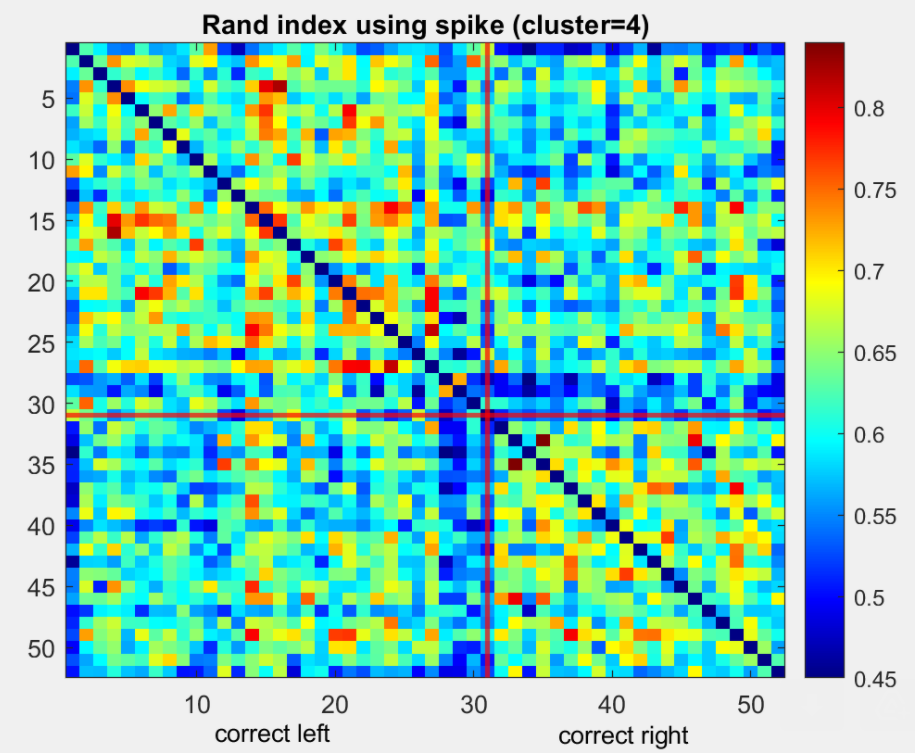}
		\end{tabular}
    		\caption{Rand index in water lick data of a mouse in one session (using four clusters) (Left: using calcium trace. Right: using estimated spike data).
    		Each element in the matrix is
the Rand index for the 2 trials of the corresponding row and column. First 31 trials: correct left trial. Last 21 trials: correct right trial. }
    \label{fig:rand_water_lick}
	\end{figure}

Figure \ref{fig:rand_water_lick} shows that clusters based on estimated spike data are more consistent between trials than clusters based on calcium trace data. Similar conclusions are obtained when the neurons are clustered to three or five clusters (appendix A.1). Because the mouse was trained before the calcium recording, individual neurons were likely have been tuned and this type of consistency across trials is expected. Therefore, the higher rand index using spike data over calcium data might indicate better performance in cluster analysis.

	\FloatBarrier

\subsection{A Fear-based contextual discrimination experiment}
In one of our previous studies \citep{johnston2020robust}, mice were trained to recognize two contexts via fear conditioning (foot shock). Meanwhile, fluorescent miniature microscopes \citep{ghosh2011miniaturized} were used to track cell populations in contextual discrimination experiments in mice's hippocampus using a genetically encoded calcium indicator (GCaMP6f) for up to 60 days. The whole experiment included four stages: habituation (mice freely exploring environment), learning (learning to freeze in a stimulus context with foot shock), extinction (no foot shock), and relearning (stimulus reinstated). 

\begin{figure}[h]
		\centering
		\begin{tabular}{c c}
			\includegraphics[scale=0.48]{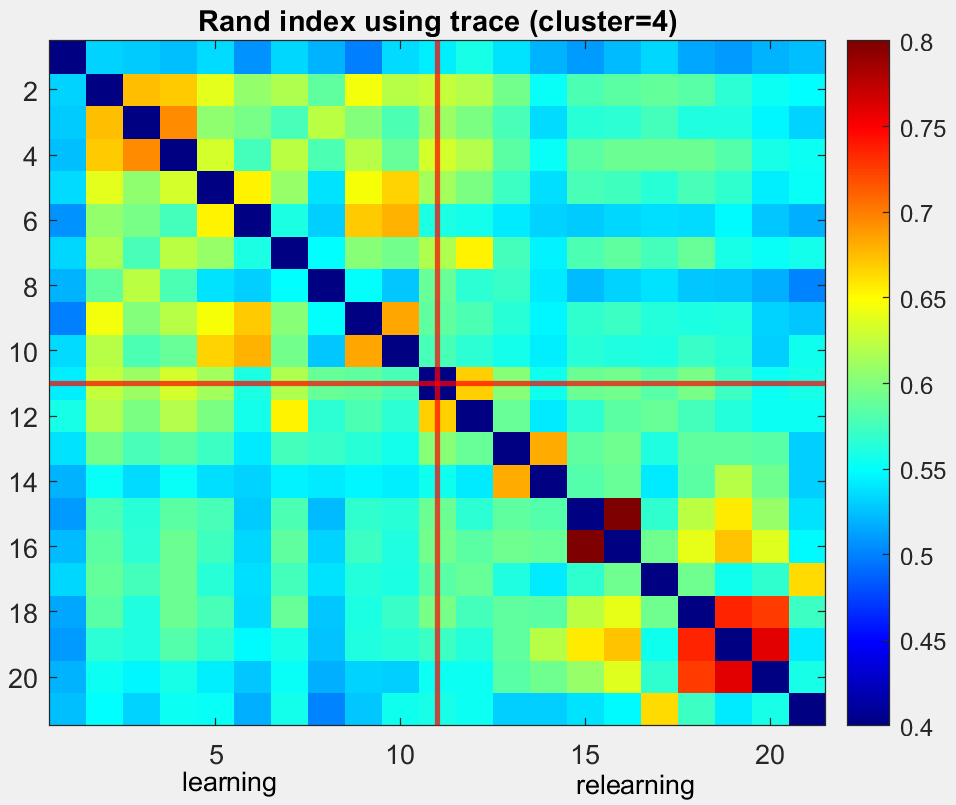}&\includegraphics[scale=0.48]{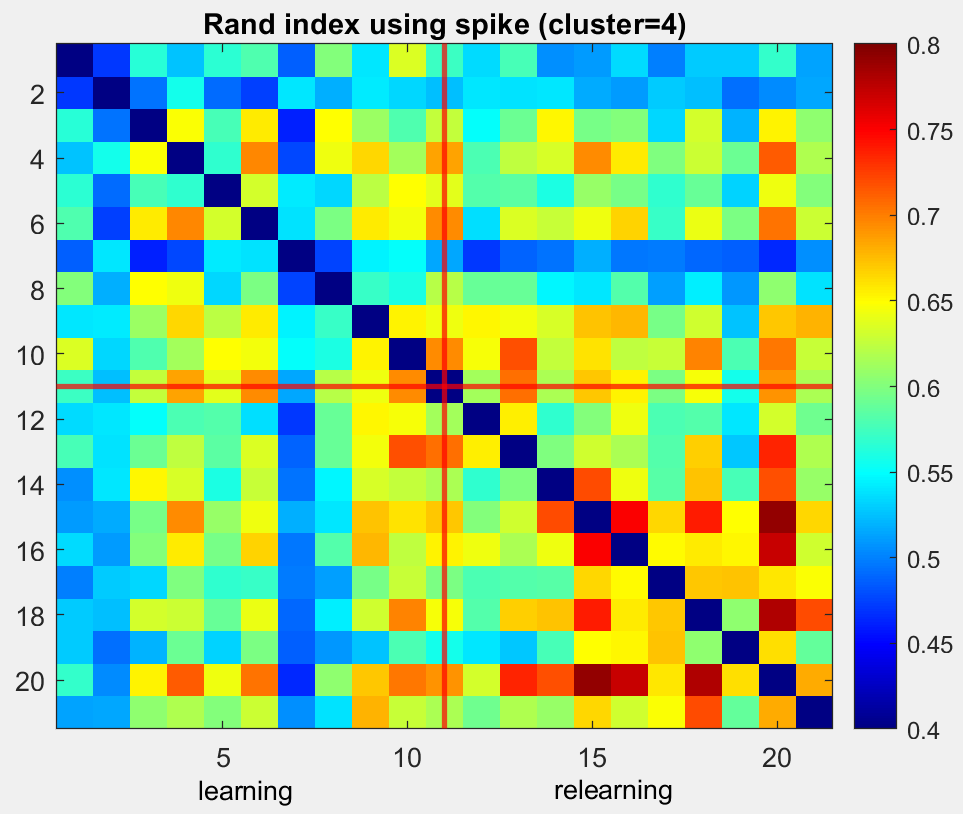}
		\end{tabular}
		\caption{Rand index in fear conditioning data of a mouse in one shock session (using four clusters). (Left: using calcium trace. Right: using estimated spike data).
		Each element in the matrix is
the Rand index for the 2 trials of the corresponding row and column. First 11 sessions: learning session. Last 10 sessions: relearning session.}
    \label{fig:rand_foot_shock}
	\end{figure}

We analyze a mouse whose 141 neurons were measured in 21 trials, with the first 11 trials in the learning stage and the last 10 trials in the relearning stage. For each trial, we clustered the neurons into four clusters using the fuzz k-means \citep{fellous2004discovering} using the calcium traces or deconvolved spike trains. The rand index based on the clustering results were computed for each pair of trials. The comparison between calcium traces and estimated spike data showed again that estimated spike trains leads to more trial-to-trial agreement in neuron clusters (Figure \ref{fig:rand_foot_shock}). Of particular interest is the higher consistency of the relearning trials relative to learning trials, which may suggest that a stable neuronal ensemble has been formed. 

The conclusion is similar when the neurons were clustered to three clusters (Figure \ref{rand_foot_3clus}). When the neurons were clustered to five clusters (Figure \ref{rand_foot_5clus}), although the rand indices for some trial pairs were lower from spike data than from calcium trace data, again, we noticed the high concordance between the clusters in the relearning trials. Due to the lack of ground truth in neuron clusters, cautions should be taken when relating results to potential misspecifications of cluster numbers. We leave further investigations to future research.
\FloatBarrier

\section{Principal Component Analysis (PCA)}

PCA is one of the most common methods in analyzing large scale neuron data \citep{chapin1999principal, cunningham2014dimensionality, churchland2012neural, harvey2012choice, kobak2016demixed}. In PCA analysis of neuron data, PCA extracts principal components that are linear combinations of individual neurons. For example, in \cite{bekolay2014spiking}, PCA was used to capture major patterns of firing within the neuronal populations in prefrontal cortex. They looked in to the first two principal components to analyze the firing patterns in motor cortex by looking at example neurons and the histogram of the loadings of neurons onto the PCs. 

Because the baseline firing rate is usually less than 1 spike/s, spike train data for most neurons are sparse. 
Thus, the conventional PCA method is not reliable. One way to handle sparsity is to consider kernels that are appropriate for sparse data. \cite{paiva2010inner} derived PCA for spike trains based on a spike train inner product (see appendix). They choose symmetric and positive definite kernels to operate on spike times. For example, the Gaussian kernel $\kappa\left(t_{m}, t_{n}\right)=\exp \left(-\frac{\left(t_{m}-t_{n}\right)^2}{\tau}\right)$ can be used to calculate the inner product of two spike times $t_m,t_n$. 

\begin{figure}[h]
		\centering
		\begin{tabular}{c c}
			\includegraphics[scale=0.5]{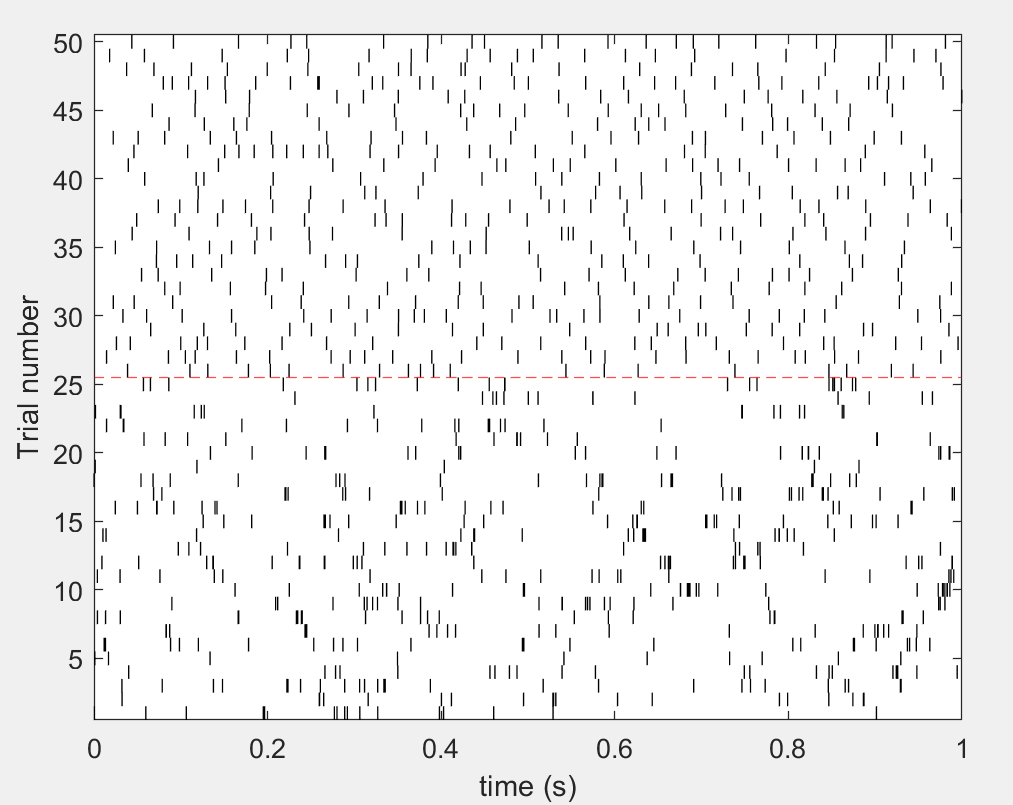}
		\end{tabular}
		\caption{The raster plot of 50 simulated spike trains (1 second with rate 20 spikes/second). Top half: 25 spike trains with shape parameter of inter-spike interval $\theta = 0.5$. Bottom half: 25 spike trains with shape parameter of inter-spike interval $\theta = 3$.}
		\label{fig: raster plot of 50 simulated spike trains}
	\end{figure}

We generate spike trains using the method described in \cite{paiva2010inner} where the spike trains are 1s long and have mean spike rate 20 spikes per second. The inter-spike interval was gamma distributed with shape parameter $\theta = 0.5$ and 3 (See Figure \ref{fig: raster plot of 50 simulated spike trains}). For each generated spike train, a calcium fluorescence trace is simulated from an  auto-regressive model \citep{vogelstein2010fast} with decay rate $\gamma=0.96$, the magnitude of each spike $s(t)=1$ and the standard deviation of noise $\sigma=0.1$ and 0.5. 

\begin{figure}[h]
		\centering
		\begin{tabular}{c c}
			\includegraphics[scale=0.4]{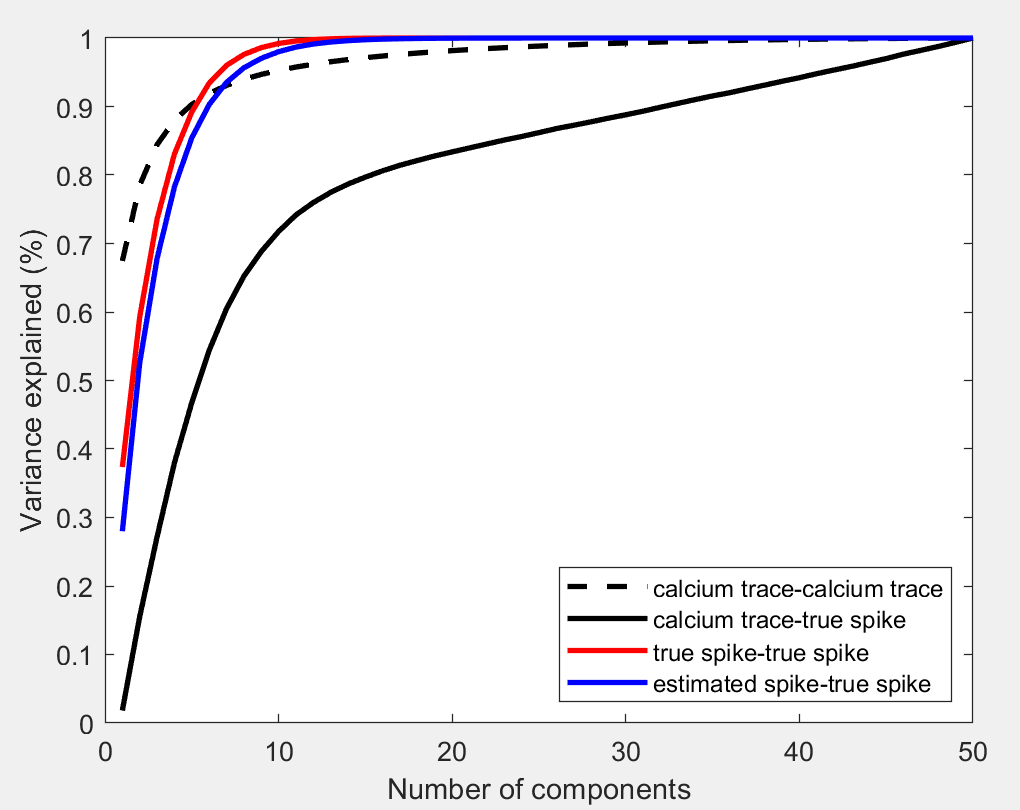}&\includegraphics[scale=0.4]{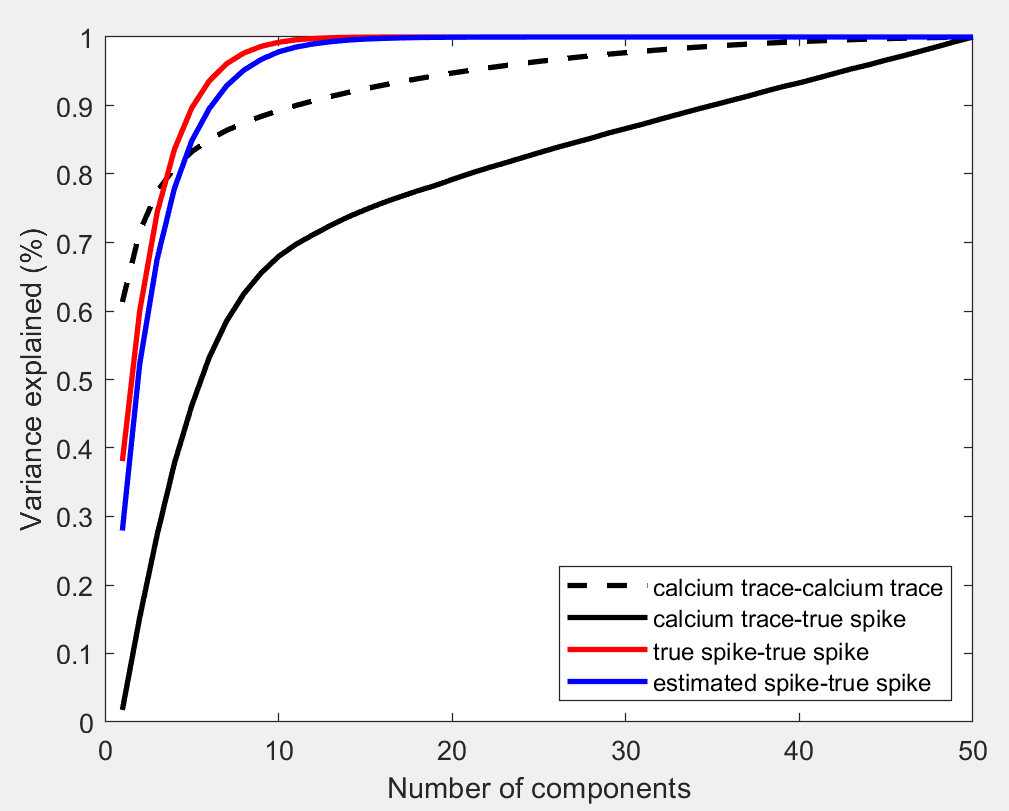} \\
		   \includegraphics[scale=0.42]{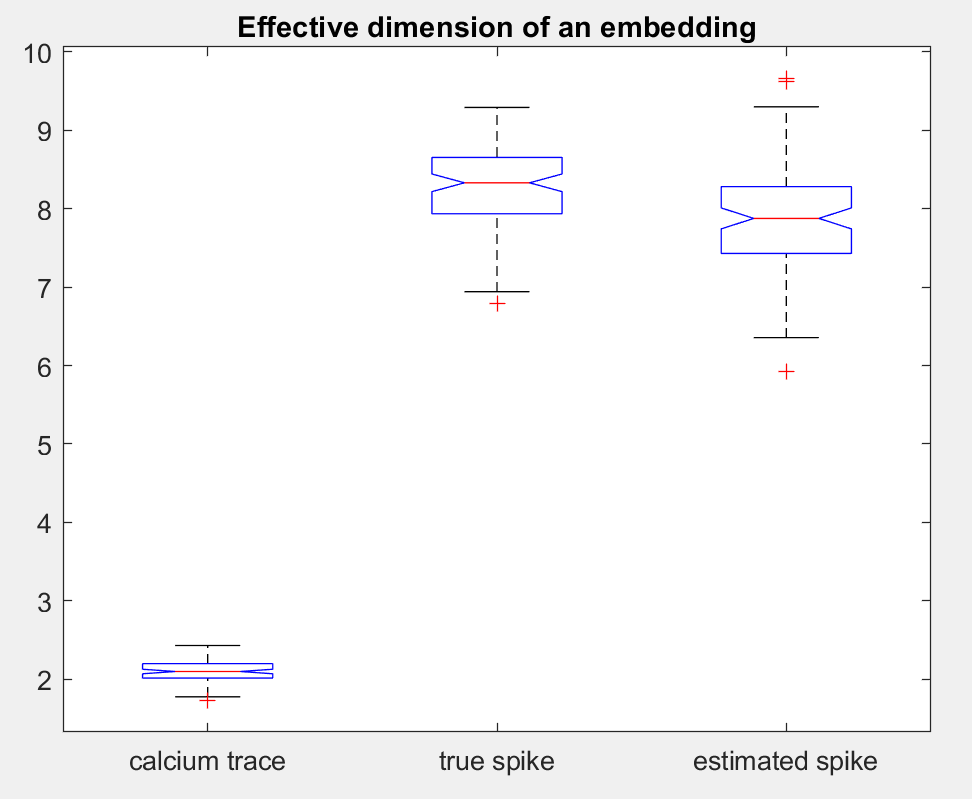}&\includegraphics[scale=0.41]{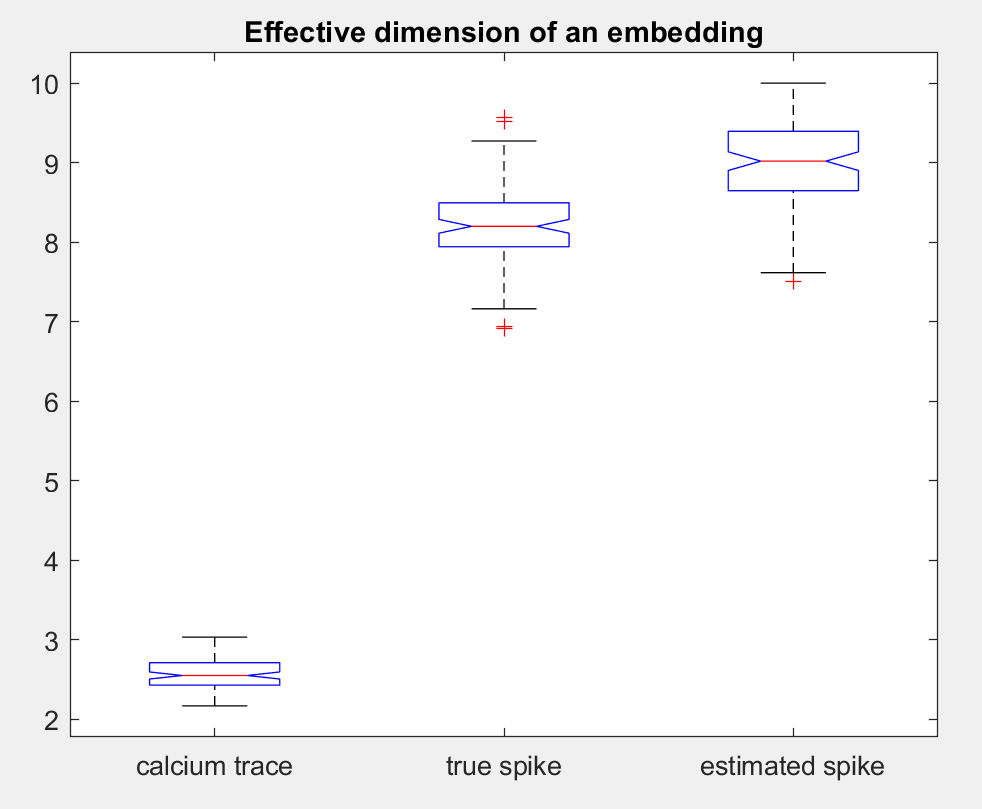}
		\end{tabular}
		\caption{Cumulative variance explained by different methods: true spike trains (red), estimated spike trains (blue), and calcium traces (black), . 
		Each curve is averaged over 100 simulated data sets. Left: noise of calcium trace $\sigma=0.1$. Right: noise of calcium trace $\sigma=0.5$. 
		Bottom left: Boxplot of the effective dimensions of calcium trace/true spike data/estimated spike data, noise of calcium trace $\sigma=0.1$. Each plot contains the result for 100 simulated data sets. 
		Bottom right: effective dimension of the calcium trace/true spike data/estimated spike data, noise of calcium trace $\sigma=0.5$. Each plot contains the result for 100 simulated data sets.
		}
		\label{fig: pca sim results}
	\end{figure}

We first compare the estimated components in explaining the variances of the true spike trains of neurons. The simulation result is averaged over 100 simulated data sets. The cumulative proportions of explained variances are shown in the upper panel of Figure \ref{fig: pca sim results}. Note that the curves based on the true spike trains (red) are very similar to those based on the estimated spike trains (blue) are quite different from those based on the calcium traces (black). The results are not surprising, as calcium traces are the noisy and integrated measurements of the true underlying spike activities. The PCA analysis using calcium traces also indicates that it requires a much large number of components to reach a certain percentage of variance. For example, when $\sigma=0.1$, it requires less than five components for the true or estimate spike train but over 15 components for the calcium traces.  


A summary metric based on the eigenvalues is the effective dimension of an embedding defined in \cite{victor1997metric}, which is also noted as the dimensionality \citep{gao2017theory}. The effective dimension index $E$ is defined as $E=\frac{\left(\sum \lambda_{i}\right)^{2}}{\sum \lambda_{i}^{2}}$ where $\lambda_i$ is the $i$-th eigenvalue of the covariance matrix of calcium traces/true spike trains/estimated spike trains between neurons. The result in the lower panel of Figure \ref{fig: pca sim results} agrees with the explained variance and shows that the dimensionality calculated from calcium traces can be quite different from those the true and estimated spike trains.

	\FloatBarrier

	\FloatBarrier

\section{Population Decoding Analysis}
Our goal in this section is to investigate whether deconvolution is helpful in population decoding. We follow a common practice that builds a predictive model using the neural activities at each time bin to predict an variable of interest\citep{berens2012fast, meyers2013neural, wei2020comparison}. 

\subsection{Data ``pre-processing"} 
As reported in \cite{wei2020comparison}, electrophysiology and estimated spike data are less informative in population decoding at each time bin. This is somewhat surprising but not totally unexpected. Two possible explanations are the integration effect of the calcium trace data \citep{wei2020comparison} and the sparsity of spike data in each time window. One support of these explanations is that the one-second filtered electrophysiology data reached higher peak decoding accuracy than the original analysis of electrophysiology data with a bin size 67ms. In fact, temporal filtering is a common practice in population decoding and choosing the appropriate filtering encourages data integration temporally, which might lead to improved decoding accuracy \cite{yates2020simple}. Here we consider two strategies to take neural firing history into consideration by applying ``pre-processing" to the sparse count data. The first method is to use cumulative counts, which is defined as the total number of spikes up to a time $t$. The second method, which is labeled with ``spike history", uses the matrix of the time series (up to time $t$) from a neuron population as input in decoding.

\subsection{Data types} Note that the spike detection framework based on AR(1) model in \ref{eqn:ar1} not only produces  estimated timestamps of spike events but also provides estimates of spike magnitudes, denoted by $s(t)$. The spike magnitude $s(t)$ was originally motivated by the number of spikes in a small time bin \cite{vogelstein2010fast}. Although  the interpretation of a positive $s(t)$ is not clear, it might still provide useful quantitative information related to spike counts and firing rates and therefore useful for population decoding. Another quantity estimated by the AR(1) model is the calcium concentration $c(t)$ in the AR(1) model. The estimated $c(t)$ can be considered as a denoised version of the observed noisy calcium trace. Interestingly, $c(t)$ can also be viewed as a filtered version of the underlying spike train. As spike train filters are often applied in various spike train analysis in neuroscience research, it is reasonable to use the estimated $c(t)$ as a candidate for population decoding. Thus, we will focus on the following four types of data: the raw calcium traces, the estimated spike trains, the estimated changes $s(t)$, and the estimated/denoised trace $c(t)$.  

\subsection{Prediction methods}
Taken together, the data types and ways of pre-processing create various combinations of decoding, as shown in Table \ref{table:2}. Note that in simulated data the ``spike train'' category includes true and estimated spike data whereas in a real study it includes spike train data from electrophysiology measurements and estimated spike train from calcium trace data.

\begin{table}[h!]
\centering
 \begin{tabular}{|c|c c c c ||} 
 \hline
  & calcium trace & estimated trace $c(t)$ & spike train& $s(t)$ \\ [0.5ex] 
 \hline\hline
 original & $\checkmark$ & $\checkmark$ & $\checkmark$ & $\checkmark$ \\ 
 \hline
 cumulative &  & & $\checkmark$ & $\checkmark$ \\ 
 \hline
 history &  & & $\checkmark$ & $\checkmark$ \\
 \hline
\end{tabular}
\caption{
Population decoding methods. The ``cumulative'' and ``history'' method for trace data are excluded due to lack of justification. }
\label{table:2}
\end{table}

In decoding with trace and cumulative count, we apply linear support vector machine (SVM) \citep{cortes1995support} with 5-fold cross validation on the training data. To assess the dependency of results on predictive models, we also used Fisher's linear discrimant analysis. Because they give similar results in the scenarios we consider, only results from SVM will be reported. In decoding with spike history, the input predictor is a matrix with time and neuron dimension. Traditional method such as SVM reshapes the matrices into vectors, causing a loss of structural information. Here we apply a sparse support matrix machine method (SSMM) \citep{zheng2018sparse}, which is a regularized binary matrix classifier that uses a $\ell_1$ penalty to ensure sparsity and a nuclear norm penalty to encourage low rank of the coefficient matrix. 

\subsection{Decoding accuracy for the water lick data}
We evaluate the usefulness of deconvolution on population coding using a subset of the data published in \cite{wei2020comparison}.
Their primary goal was to compare spike and imaging data for measuring neuronal activities. They recorded neural activities in anterior lateral motor cortex using electrophysiology signals and calcium imaging. Although the electrophysiologyand calcium imaging recordings were separate, neuron populations of matched depth were measured from the same delayed response task. Each 5 second long recording was composed of three epochs: sample epoch (mice presented with a vertical pole), delay epoch (the pole was removed), response epoch (mice cued to give a response). The duration of sample and delay epoch was 2.6 s. In electrophysiology, the sample epoch was 1.3 s while in calcium imaging was 1.2 s. Here we analyzed the calcium imaging data for 1,493 neurons from 4 mice with adeno-associated virus expressing GCaMP6s and electrophysiology data of 720 neurons from 19 mice. To match the temporal resolution of calcium trace data, the spike train data from electrophysiology recordings were temporally subsampled. Note that the data we downloaded from \emph{https://doi.org/10.6084/m9.figshare.12786296.v1} has been pre-processed. We also followed several other strategies used in \cite{wei2020comparison}, such as resampling trials to cope with the limited number of trials, resampling the same number of neurons when comparing different data types, using cross validation to choose tuning parameters ($60\%$ data for training and $40\%$ data for testing). The full description of the experiments and data processing is available from \cite{wei2020comparison} and further backgrounds about the experimental design can be found in other articles of the group \citep{li2015motor, wei2019orderly}.  

\begin{figure}[h]
		\centering
		\begin{tabular}{c }
			\includegraphics[scale=0.4]{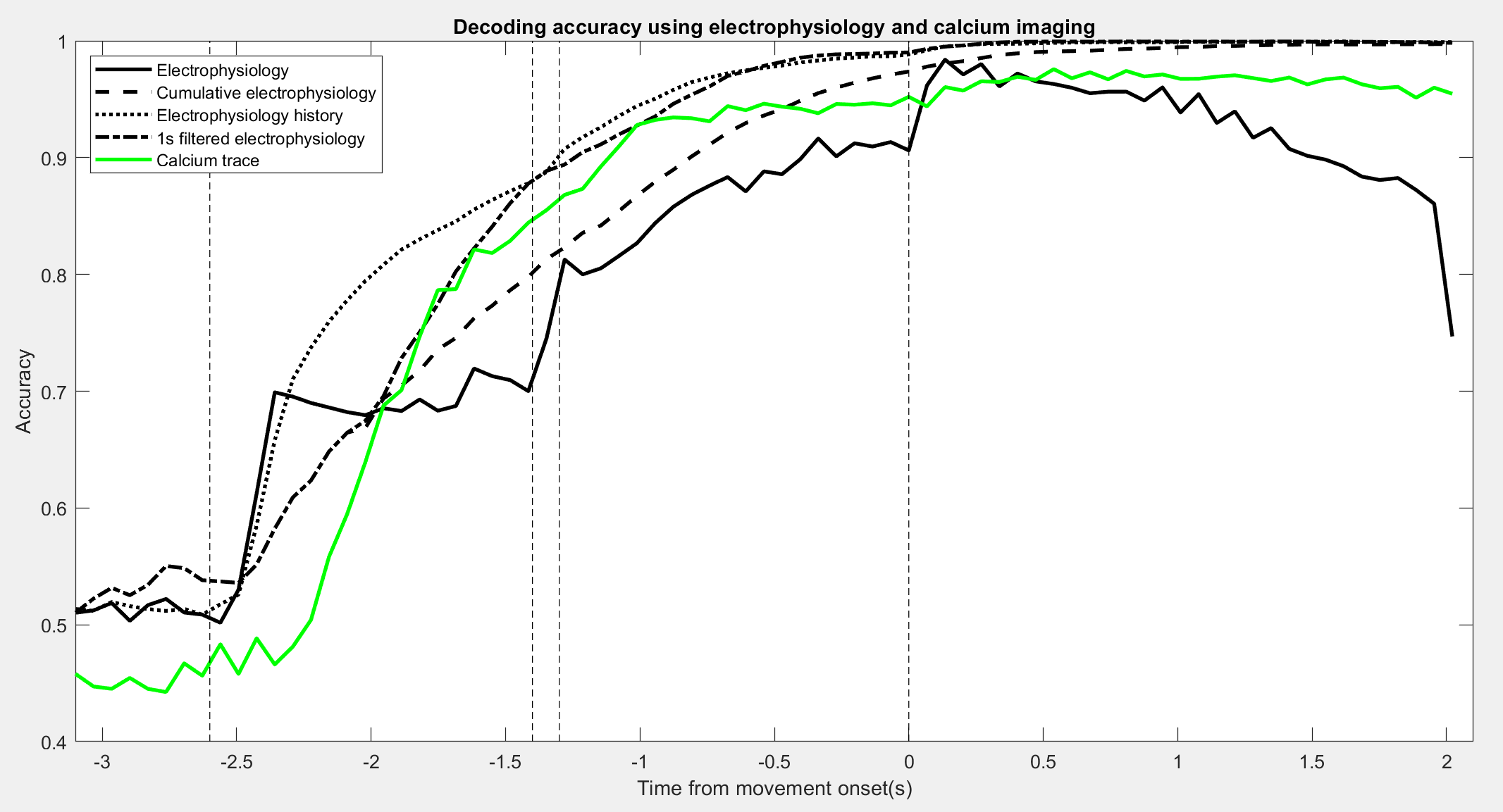}\\
			\includegraphics[scale=0.4]{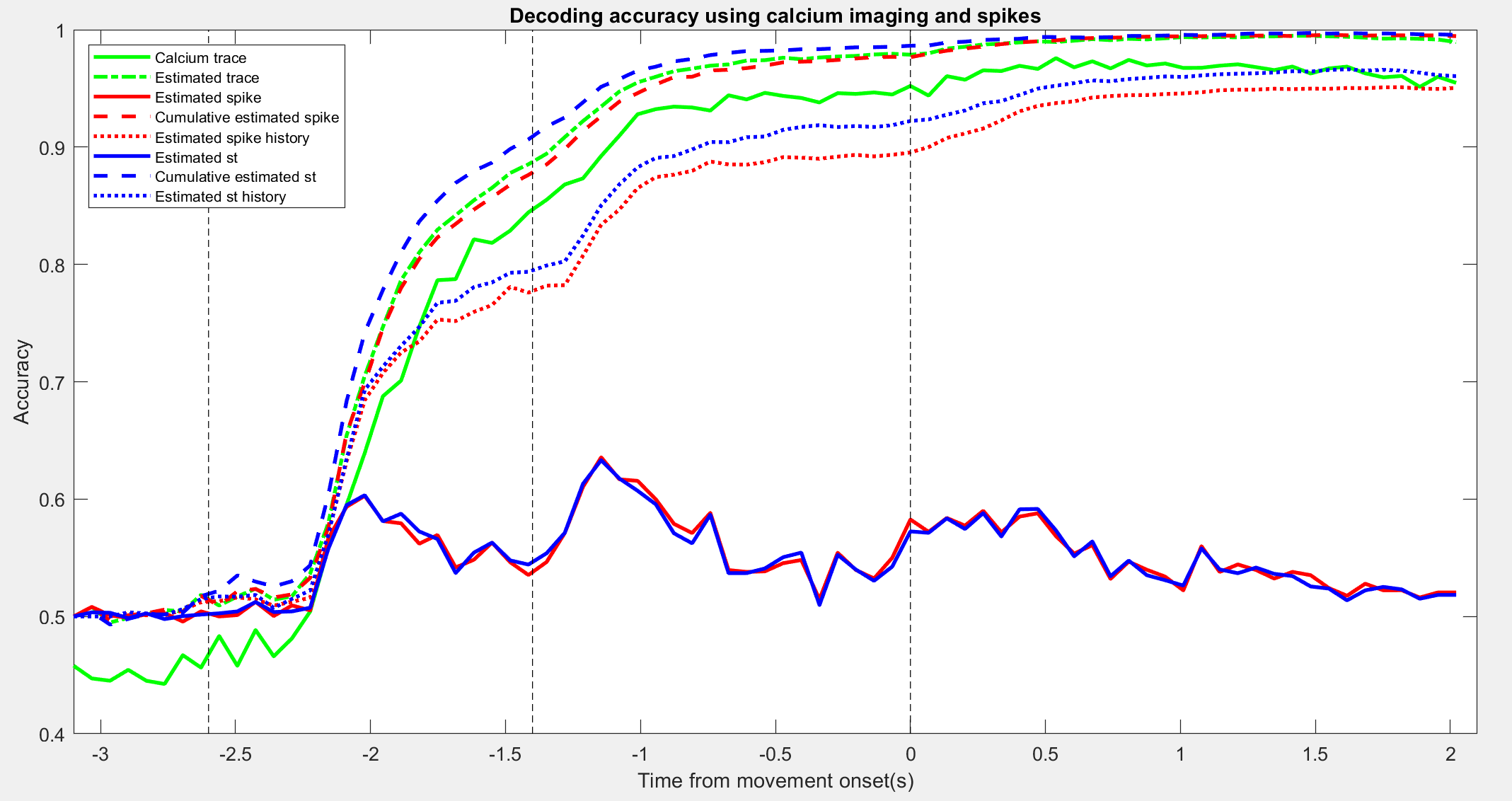}
		\end{tabular}
		\caption{Decoding results in water lick data.
		Top: decoding accuracy using electrophysiology and calcium imaging data. Black lines: electrophysiology. Green lines: calcium trace.
		vertical dashed lines from left to right: the start time of sample epoch, delay epoch of calcium imaging data, delay epoch of electrophysiology, and response epoch.
		Bottom: decoding accuracy using calcium imaging data and spikes. Green lines: calcium trace. Red lines: Estimated spikes from calcium trace. Blue lines: Estimated spike magnitude from calcium trace. Decoding results are averaged over 100 subsamples. In each subsample, 50 neurons are randomly selected and 500 trials are sampled from each neuron.  vertical dashed lines from left to right: the start time of sample epoch, delay epoch and response epoch.}
		\label{fig:population_decoding_water_lick}
	\end{figure}

Figure \ref{fig:population_decoding_water_lick} visualizes the accuracy, which is defined as the proportion of correct predictions in testing data, of the prediction methods in Table \ref{table:2}. The top panels replicated the difference between the decodability of the calcium trace and the electrophysiology recordings in \cite{wei2020comparison}. Although electrophysiology data showed earlier latency (defined as the time reaching $\%70$ decoding accuracy), surprisingly, it had lower decoding accuracy than calcium trace data, which are a noisy measurement of neural activities. This difference might be due to the integration effect of the calcium trace data, as a calcium transient is characterized by a rapid rise but slow decay. This hypothesis is supported by the improved decoding accuracy of the 1-second filtered electrophysiology data in most time bins. However, the 1-second filtered electrophysiology data lost the early latency. As a comparison, electrophysiology history, a new method considered here, outperforms the calcium trace data in all time bins. Thus, when used appropriately, the less noisy spike train data from electrophysiology recordings do have higher population decodability. 

The lower panel of Figure \ref{fig:population_decoding_water_lick} focused on the effects of deconvolving calcium traces. Consistent with \cite{wei2020comparison}, the estimated spike data has lower decodability than both calcium traces and electrophysiology data. The magnitude of changes, i.e., $s(t)$ has a similar accuracy rate. However, their cumulative and history have much higher accuracy, with their cumulative outperforming the calcium trace data. It is worthy pointing out that the estimated calcium concentrations (``estimated calcium trace"), which is a denoised version of the calcium traces, also demonstrates higher decoding accuracy than the original calcium traces. These results indicate that the decodability of a neuron population can be improved by appropriate analysis of the deconvolved data.

	\FloatBarrier

\section{Discussion}
In this article, we compared calcium traces and the estimated spike data in three analyses: cluster analysis, PCA, and decoding. Using simulated data, we showed that deconvolution leads to better estimates of neural clusters and principal components than do calcium traces. In our analysis of a real study, the estimated spike trains produced neural clusters with higher trial-to-trial concordance and, when analyzed in appropriate ways, they also give more accurate results in population decoding. 

For computational efficiency, we chose a recently published $\ell_0$ penalized method \citep{jewell2018exact} to estimate spike trains from calcium traces. Because different deconvolution methods have varying accuracy in spike estimation it is expected that our comparison depends on the choice of the choice of deconvolution method, as reported in \citep{evans2019use}. On the other hand, given the consistent observations in our study and previous studies \citep{evans2019use, wei2020comparison}, it is reasonable to assume that the difference between choosing calcium trace and estimated spike data is mainly due to the nature of the data, such as quantitative vs binary, continuous vs sparse, and delayed/integrated vs instantaneous. 

The results also likely vary from the specific analysis methods chosen, such as choice of clustering methods, choice of kernel in the kernel PCA for spike data, and decoding methods. For example, when using the cumulative spike data for population decoding, we use the whole time course no later than the current time window. In a recent work \cite{yates2020simple} showed that population accuracy varied across window size. In \cite{tu2020efficient, park2014encoding}, a temporal filter is applied to a calcium transient feature to encourage integrating neuronal activities to improve decoding accuracy. While adopting sophisticated decoding algorithms might improve the prediction accuracy for both calcium traces and the deconvolved spike data, we expect that the main conclusion remains.



\newpage
\begin{appendix}
\section{Fuzzy k-means}\label{appA}
In the standard K-means algorithm, each data point is assigned to one cluster. 
The fuzzy k-means algorithm (FKM) was first developed in \cite{dunn1973fuzzy} and improved in \citep{bezdek1984fcm,bezdek2013pattern}. It assigns each data point a probability of belonging to each cluster instead. Suppose we have the data $Y=\left\{y_{1}, y_{2}, \ldots, y_{N}\right\} \subset \mathbb{R}^{p}$. FKM is based on minimizing the objective function:
$$
J_{f}(U, C)=\sum_{i=1}^{N} \sum_{j=1}^{K} u_{i j}^{f}\left\|{y}_{i}-{c}_{j}\right\|^{2}
$$
where $U$ is a fuzzy partition of the $N$ data points $y_{i}$ with $K$ cluster centers $C=\{{c}_{1},...c_k\} \subset \mathbb{R}^{p}$. The numbers $u_{i j}$ are the degree of membership of data point $i$ to cluster $j$ (the fuzzy partition) and $f>1$ is the fuzziness factor. For most data, $1.5\le f<3$ gives good results \citep{bezdek1984fcm}. In the simulation data used in \cite{fellous2004discovering}, $f$ is chosen as 2 and it gives robust results. The membership matrix satisfies $\sum\limits_{j=1}^K u_{ij}=1$. The fuzzy partition and the cluster centers were computed iteratively. 
At each iteration, the membership matrix and cluster centers were updated as:
$$
u_{i j}=\frac{1}{\sum_{k=1}^{K}\left(\frac{d_{i j}}{d_{i k}}\right)^{\frac{2}{f-1}}}
$$
and
$$
{c}_{j}=\frac{\sum_{i=1}^{N} u_{i j}^{f} y_{i}}{\sum_{i=1}^{N} u_{i j}^{f}}
$$
where
$$
d^2_{i j}=\left\|y_{i}-{c}_{j}\right\|^2
$$
The iteration stopped when the norm of two consecutive membership matrix $\left\|U^{new}-U^{old}\right\|^2$ was smaller than a termination criterion $\epsilon$. 

\newpage
\subsection{Clustering results with different number of clusters in simulation and real data}

\begin{itemize}
\item Rand index and mutual information in simulation data (four clusters)
\end{itemize}
	\begin{figure}[h]
		\centering
		\begin{tabular}{c c}
			\includegraphics[scale=0.45]{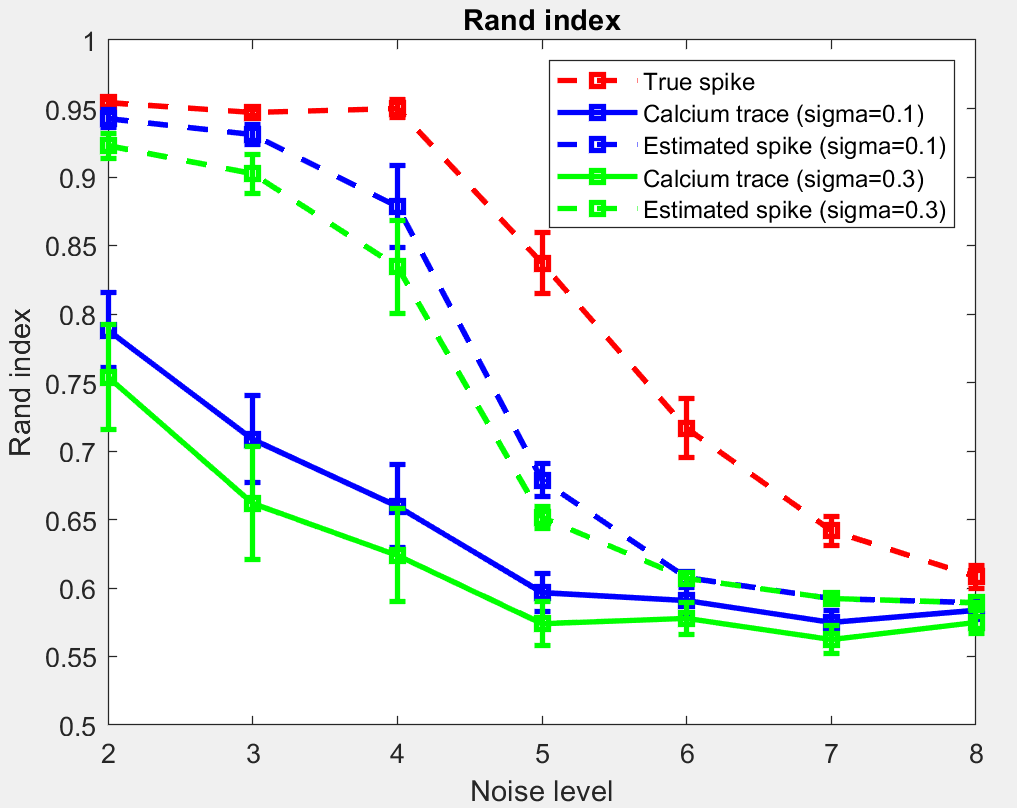}&\includegraphics[scale=0.45]{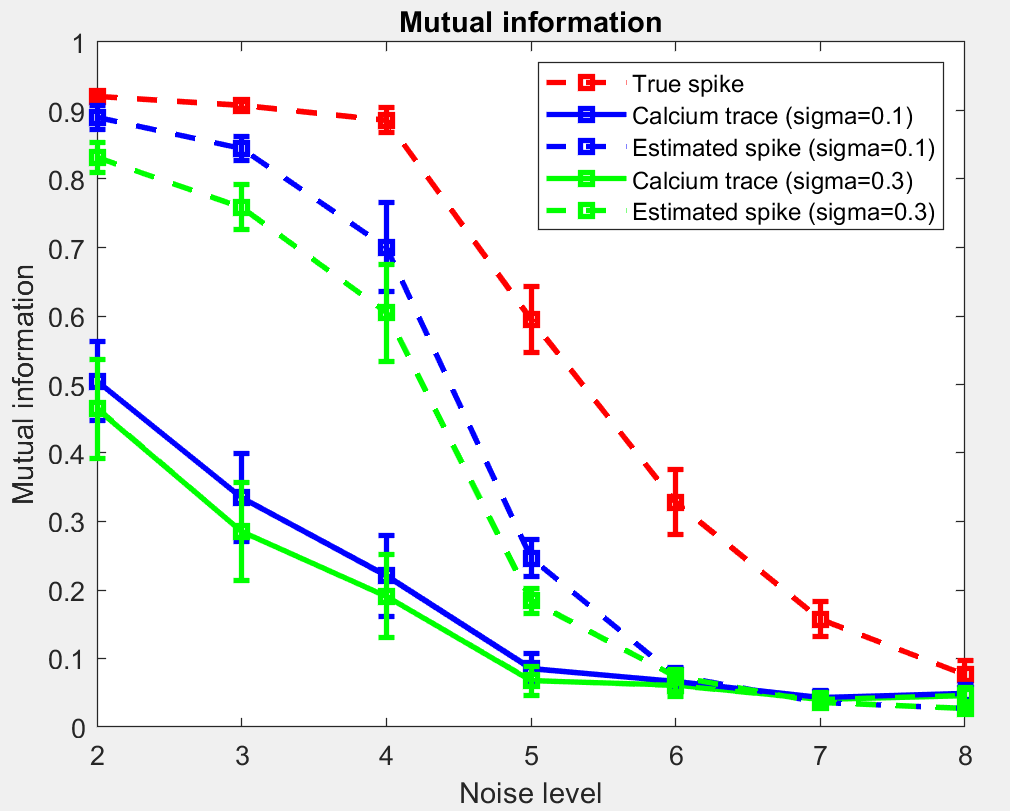}
		\end{tabular}
		\caption{Clustering results for simulation data over different noise levels with $G=3$ groups, number of extra spikes $X=2,3,4,8,11,15,20$ and time of jittering $J=1,3,5,10,15,20,30$ ms. The noise level ranges from 2 $(J=3, X=2)$ to 8 $(J=30, X=20)$. Each point is averaged over 30 datasets and 100 simulations on the same noise level. Left: rand index. Right: mutual index. Vertical bars are the corresponding 95\% confidence intervals over 30 datasets in each simulation setting.}
		\label{clus_sim_4clus}
	\end{figure}

\begin{itemize}
\item Rand index and mutual information in simulation data (five clusters)
\end{itemize}
	\begin{figure}[h]
		\centering
		\begin{tabular}{c c}
			\includegraphics[scale=0.45]{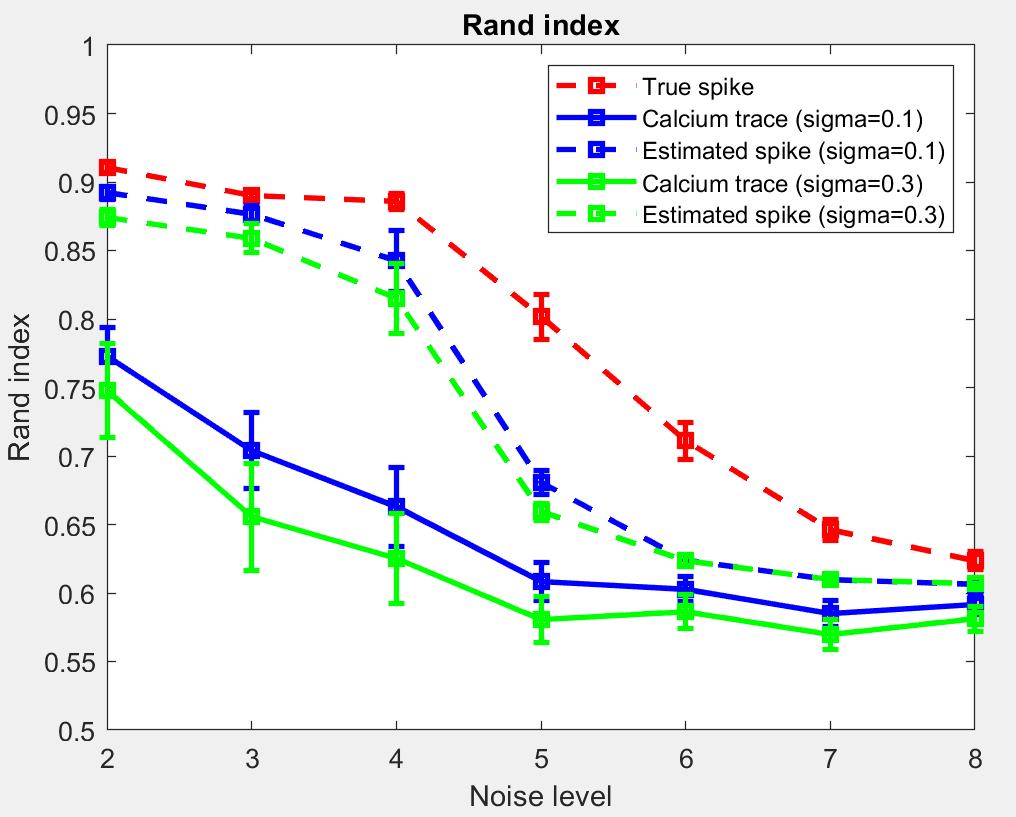}&\includegraphics[scale=0.45]{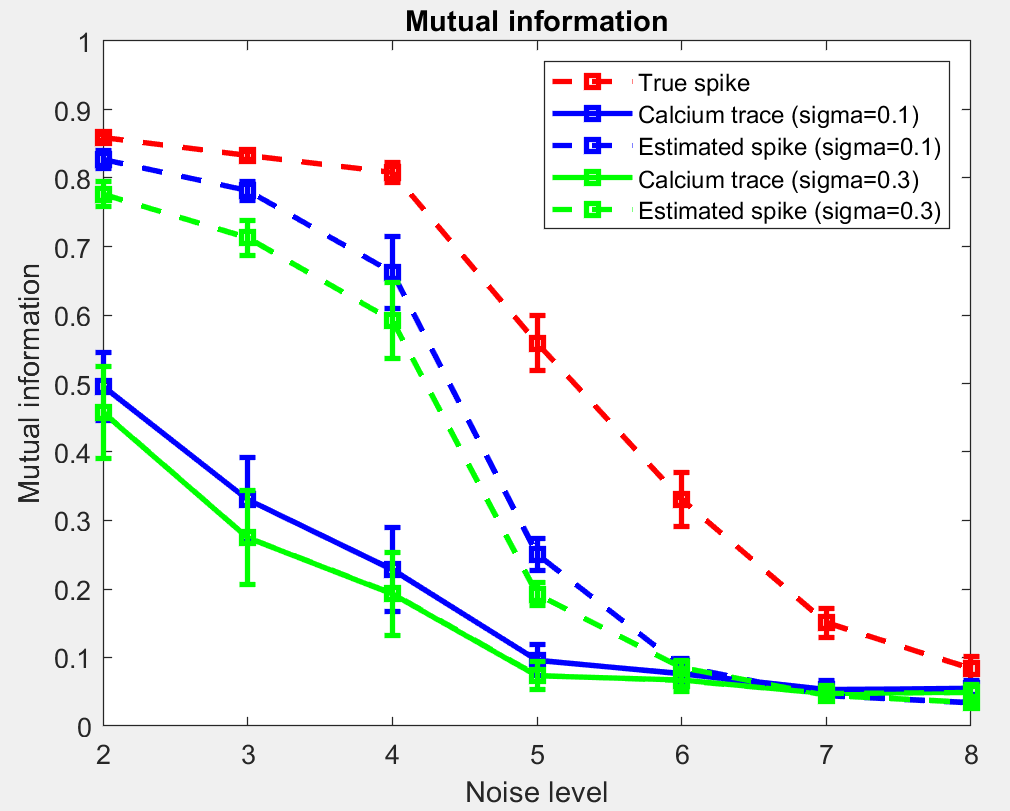}
		\end{tabular}
		\caption{Clustering results for simulation data over different noise levels with $G=3$ groups, number of extra spikes $X=2,3,4,8,11,15,20$ and time of jittering $J=1,3,5,10,15,20,30$ ms. The noise level ranges from 2 $(J=3, X=2)$ to 8 $(J=30, X=20)$. Each point is averaged over 30 datasets and 100 simulations on the same noise level. Left: rand index. Right: mutual index. Vertical bars are the corresponding 95\% confidence intervals over 30 datasets in each simulation setting.}
		\label{clus_sim_5clus}
	\end{figure}
 
\begin{itemize}
\item Rand index in water lick data (three clusters)
\end{itemize}
	\begin{figure}[h]
		\centering
		\begin{tabular}{c c}
			\includegraphics[scale=0.475]{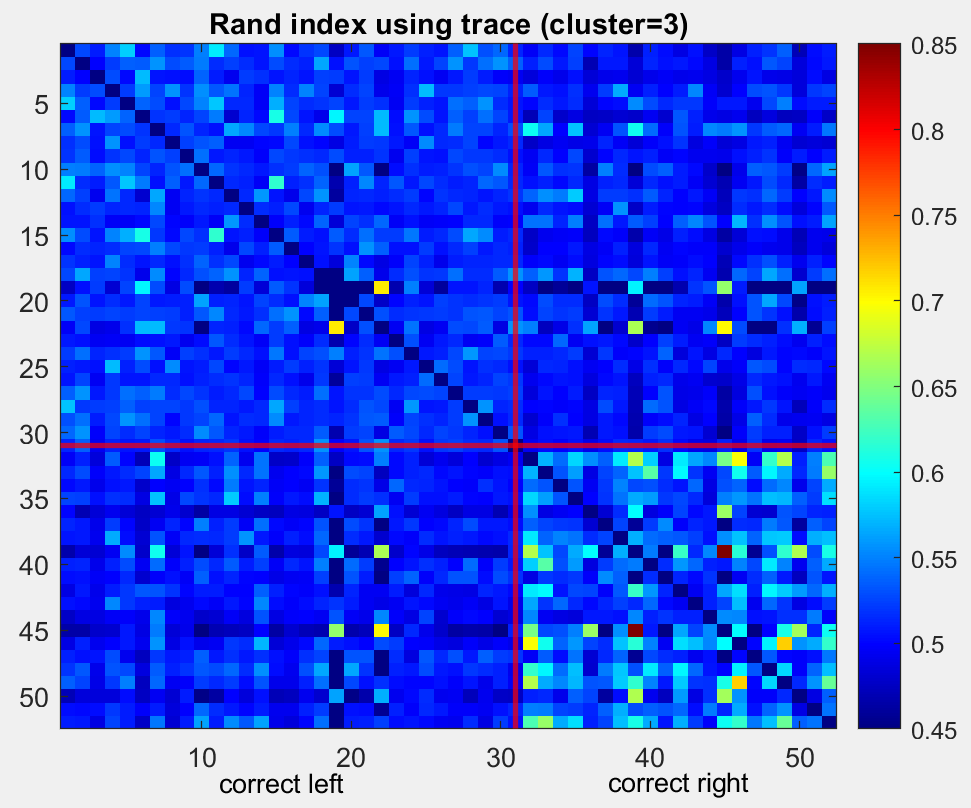}&\includegraphics[scale=0.47]{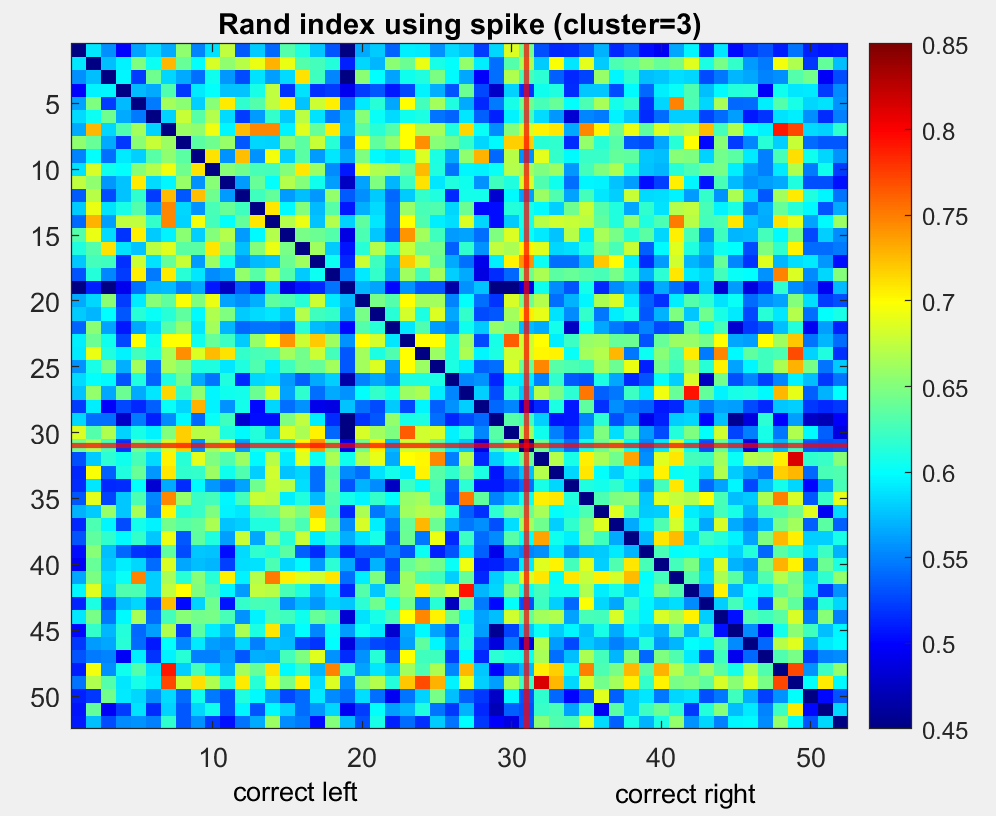}
		\end{tabular}
    		\caption{Rand index in water lick data of a mouse in one session (using three clusters) (Left: using calcium trace. Right: using estimated spike data).
    		Each element in the matrix is
the Rand index for the 2 trials of the corresponding row and column. First 31 trials: correct left trial. Last 21 trials: correct right trial. }
\label{rand_water_3clus}
	\end{figure}
	\FloatBarrier
	
	\begin{itemize}
\item Rand index in water lick data (five clusters)
\end{itemize}
	\begin{figure}[h]
		\centering
		\begin{tabular}{c c}
			\includegraphics[scale=0.47]{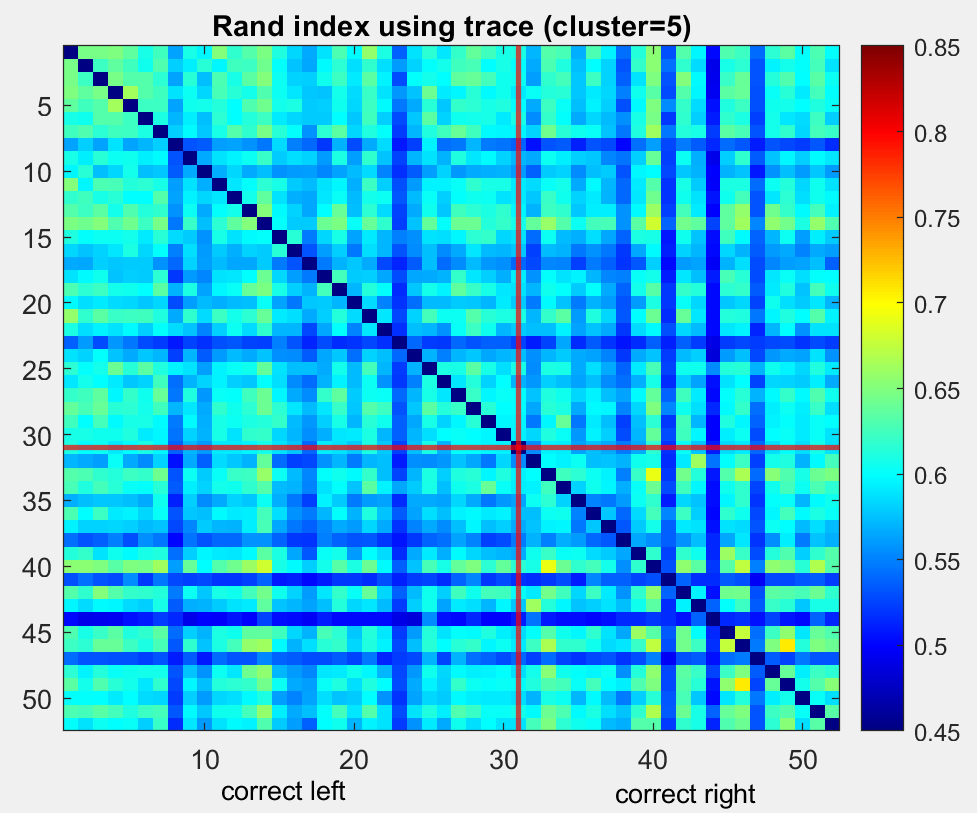}&\includegraphics[scale=0.47]{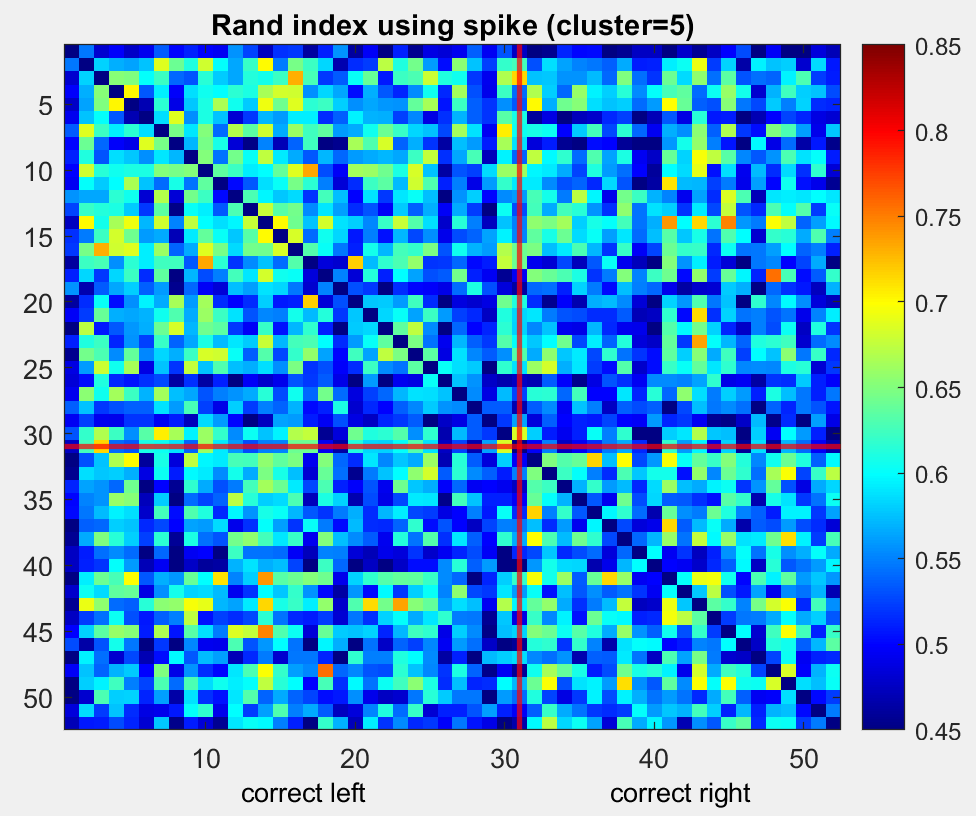}
		\end{tabular}
    		\caption{Rand index in water lick data of a mouse in one session (using five clusters) (Left: using calcium trace. Right: using estimated spike data).
    		Each element in the matrix is
the Rand index for the 2 trials of the corresponding row and column. First 31 trials: correct left trial. Last 21 trials: correct right trial. }
	\end{figure}
	\label{rand_water_5clus}
	\FloatBarrier
	
\begin{itemize}
\item Rand index in fear conditioning data (three clusters)
\end{itemize}
	\begin{figure}[h]
		\centering
		\begin{tabular}{c c}
			\includegraphics[scale=0.47]{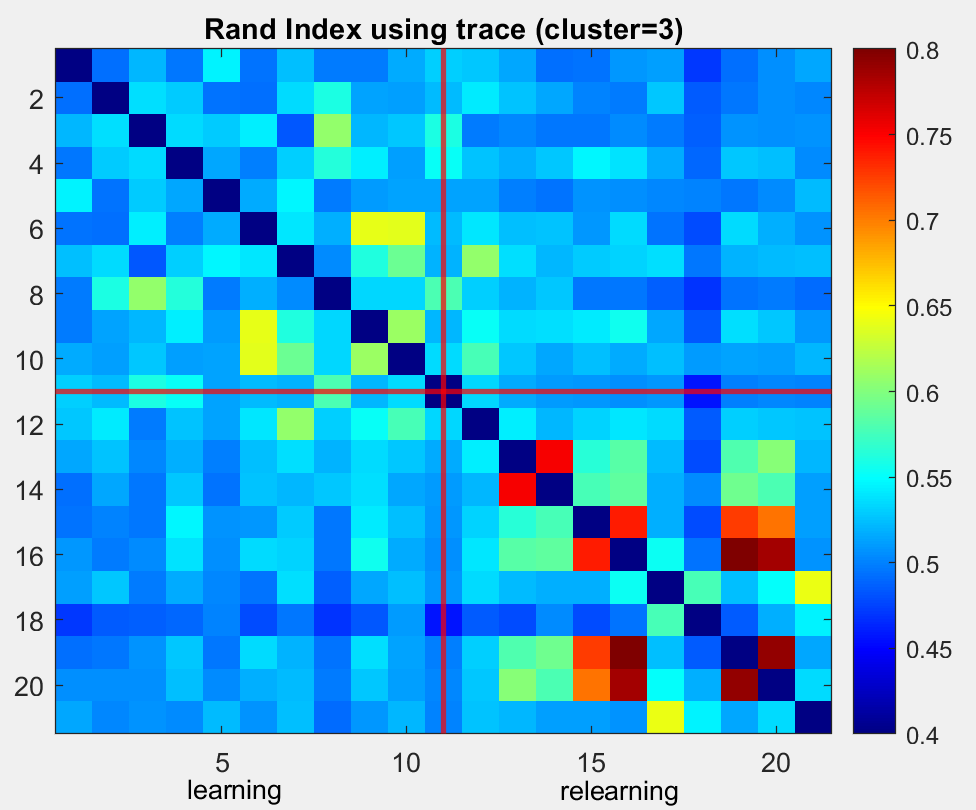}&\includegraphics[scale=0.47]{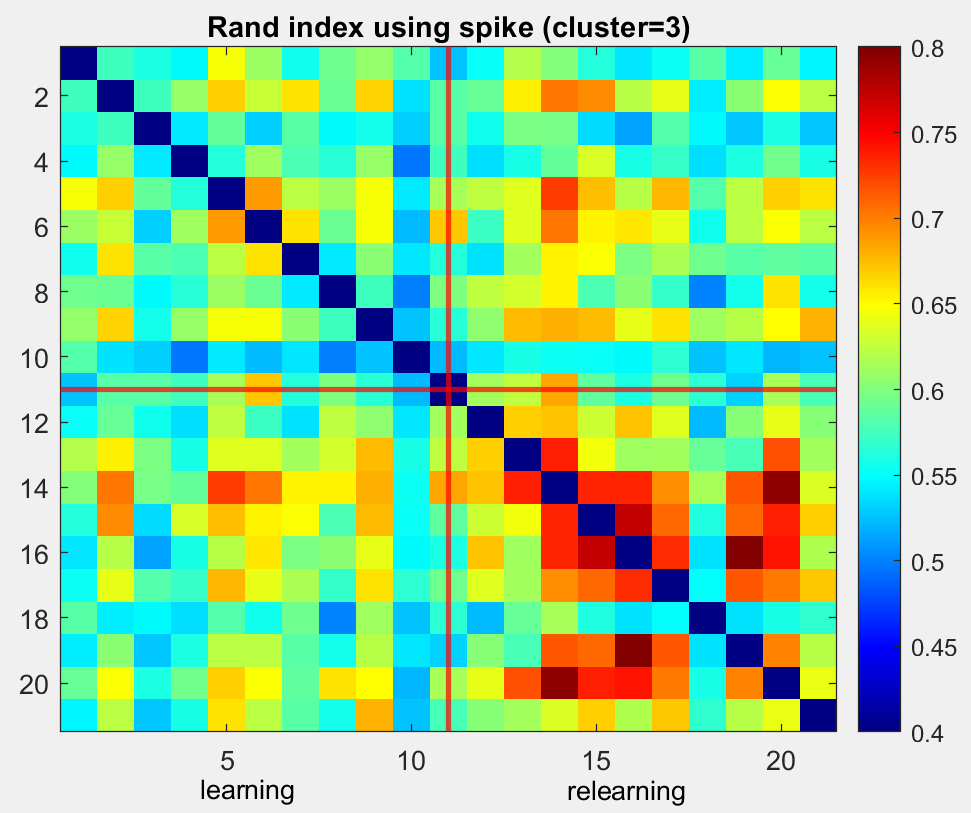}
		\end{tabular}
    		\caption{Rand index in fear conditioning data of a mouse in one shock session. (Left: using calcium trace. Right: using estimated spike data).
		Each element in the matrix is
the Rand index for the 2 trials of the corresponding row and column. First 11 sessions: learning session. Last 10 sessions: relearning session. }
\label{rand_foot_3clus}
	\end{figure}
	\FloatBarrier
	
	\begin{itemize}
\item Rand index in fear conditioning data (5 clusters)
\end{itemize}
	\begin{figure}[h]
		\centering
		\begin{tabular}{c c}
			\includegraphics[scale=0.47]{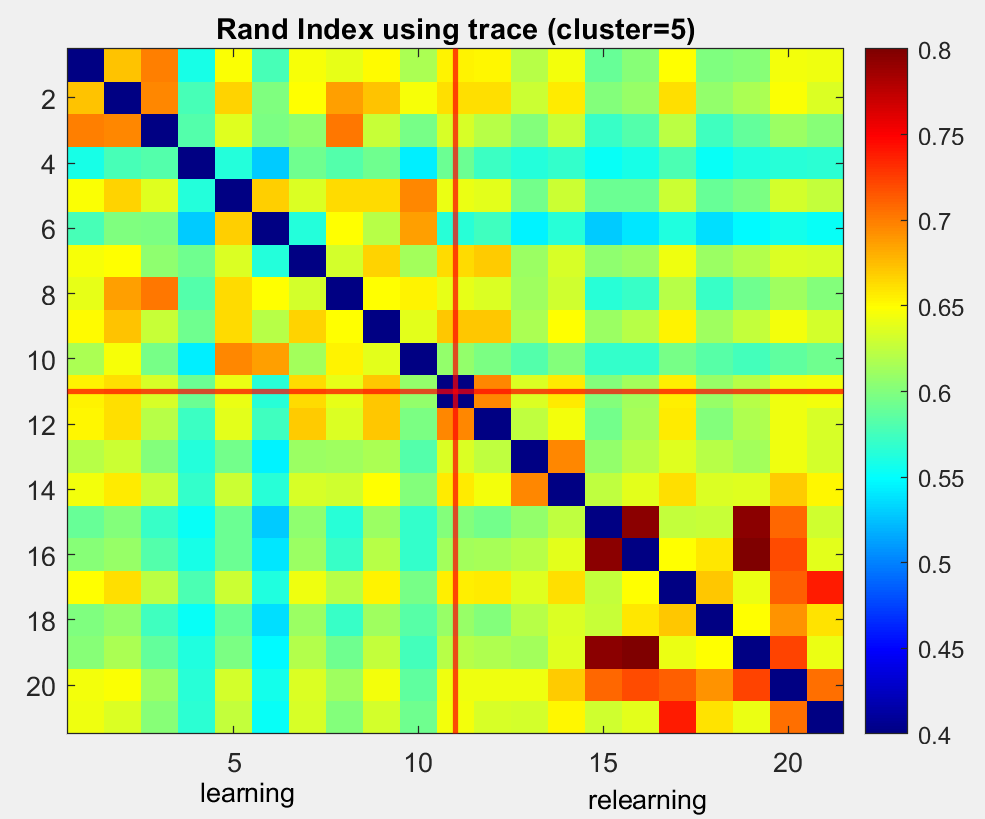}&\includegraphics[scale=0.47]{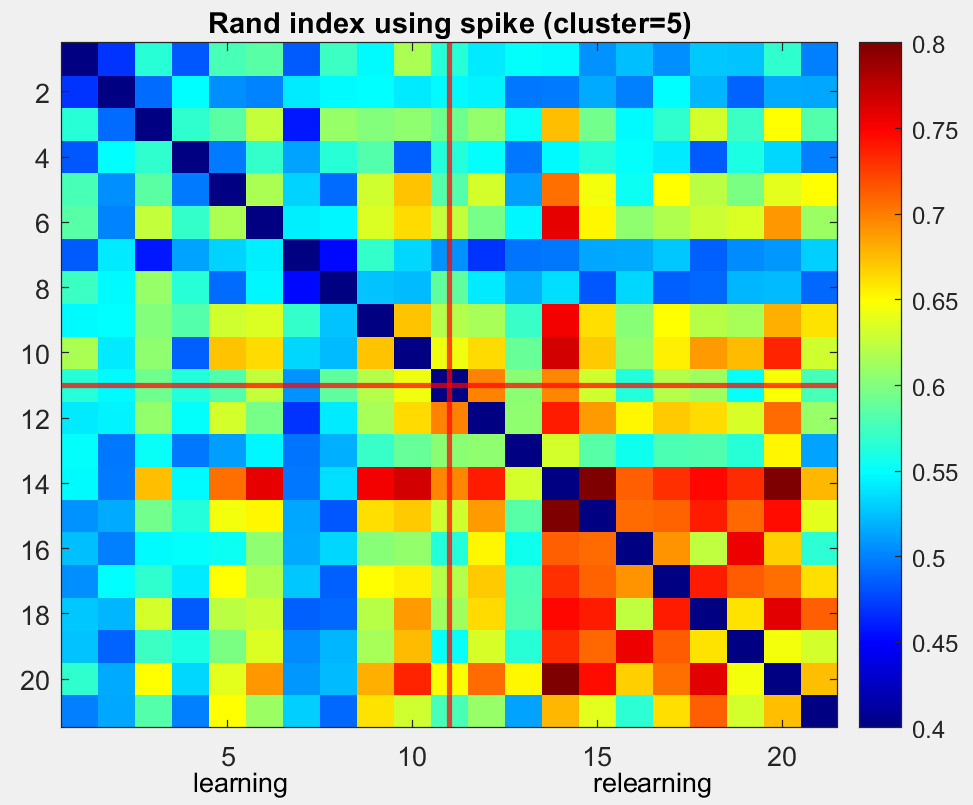}
		\end{tabular}
    		\caption{Rand index in fear conditioning data of a mouse in one shock session. (Left: using calcium trace. Right: using estimated spike data).
		Each element in the matrix is
the Rand index for the 2 trials of the corresponding row and column. First 11 sessions: learning session. Last 10 sessions: relearning session. }
\label{rand_foot_5clus}
	\end{figure}
	\FloatBarrier

\section{Spike train PCA}\label{appB}
A spike train is a binary time series usually with a ``0'' for``no-spike'' and ``1'' for an action potential. The information in a spike train is also often represented using the ordered spike times. For example, consider a spike train $s_i$ $(1\le i\le N)$ with $N_i$ spikes. We can use the set of spike times $t_1^i < t_2^i < \cdots < t_{N_i}^i$ to represent the spike. Thus, the kernel function between two spike trains should applicable to two sets, rather than two vectors. \citep{carnell2005linear} suggested compare all the spike times of one spike train to all the spike times of the other spike train, i.e.,  
\begin{equation*}
\begin{aligned}
K\left(s_{i}, s_{j}\right) 
&=\sum_{m=1}^{N_{i}} \sum_{n=1}^{N_{j}} \kappa\left(t_{m}^{i}, t_{n}^{j}\right)
\end{aligned}
\end{equation*}
where $\kappa$ is a symmetric, shift-invariant, and positive definite kernel. In this article, we use the Gaussian kernel $\kappa(t_{m}^i, t_{n}^j)=\exp \{-(t_{m}^i-t_{n}^j)^2/\tau\}$. 

The above spike train inner product has been adopted by other researchers \citep{carnell2005linear, schrauwen2007linking, paiva2009reproducing}. Recall that the components in the conventional PCA are based on the sample variance-covariance or correlation matrix, which is computed using centered data. To conduct kernel PCA using the spike train inner product, one also needs a similar centering process \citep{paiva2010inner}, which can be shown equivalent to conduct the double-centering. Specifically, let $\mathbf{P}$ denote the inner product matrix of the spike trains, where $\mathbf{P}_{ij} = K(s_i,s_j)$. Then the double centered inner product is  $\tilde{\mathbf{P}}$ is
$$
\begin{aligned}
\tilde{\mathbf{P}}
&=(I-\frac{\mathbf{1}_{N} \mathbf{1}_{N}^{T}}{N})\mathbf{P}(I-\frac{\mathbf{1}_{N} \mathbf{1}_{N}^{T}}{N})
\end{aligned}
$$
where $\mathbf{1}_{N}$ is the $N \times 1$ vector with all ones. After obtaining the centered inner product matrix, eigen decomposition is performed on $\tilde{\mathbf{P}}$ to compute the eigenvectors and corresponding eigenvalues.

\section{Sparse support matrix machine}\label{appC}
Sparse support matrix machine (SSMM) is a regularized binary matrix classifier to predict the class of the input predictor matrices.
Given a set of samples $\left\{\mathbf{X}_{i}, y_{i}\right\}_{i=1}^{n}, \mathbf{X}_{i} \in \mathbb{R}^{p \times q}$ is the $i_{t h}$
predictor matrix and $y_{i} \in\{1,-1\}$ is its corresponding label.
The proposed SSMM \citep{zheng2018sparse} is based on support vector machine \citep{cortes1995support} and it combines a hinge
loss with a new regularization on the regression matrix $\mathbf{W}$. 
The objective function of SSMM method is presented as:
$$
\underset{\mathbf{W}, b}{\arg \min } \left\{\lambda\|\mathbf{W}\|_{1}+\tau\|\mathbf{W}\|_{*}+\sum_{i=1}^{n}\left\{1-y_{i}\left[\operatorname{tr}\left(\mathbf{W}^{T} \mathbf{X}_{i}\right)+b\right]\right\}_{+} \right\}
$$
where $\mathbf{X}_{i}, \mathbf{W} \in \mathbb{R}^{p \times q}$. It incorporates the hinge loss, $\ell_1$ norm $\|\mathbf{W}\|_{1}$ and nuclear norm $\|\mathbf{W}\|_{*}$ on regression matrix $\mathbf{W}$ for matrix classification. The $\ell_{1}$ norm controls the sparsity of $\mathbf{W}$ and the nuclear norm  encourages $\mathbf{W}$ to be low-rank. For the tuning parameters $\tau$ and $\lambda$, it has been reported that non-zero values often give better results than zero \citep{zheng2018sparse}. We tried different combinations of the two parameters and compare the decoding accuracy on simulated and real data with all time points (results omitted). 

An efficient algorithm to solve the optimization problem is presented in \cite{zheng2018sparse}, where they smooth the loss function by using generalized smooth hinge loss with Lipschitz-continuous gradient. The computational cost is $\mathcal{O}\left(n^{2} pq\right)$.

\end{appendix}

\newpage

\bibliographystyle{unsrtnat}  
\bibliography{references}  

\end{document}